\documentclass[floats,floatfix,showpacs,amssymb,prd,twocolumn,superscriptaddress,nofootinbib,longbibliography,reprint]{revtex4-1}

\usepackage{amssymb,amsmath,verbatim,mathtools,needspace,enumitem,etoolbox,graphicx,physics,microtype,afterpage,bigints,gensymb,tabularx}

\usepackage{multirow}
\usepackage[dvipsnames, usenames]{xcolor}
\definecolor{linkcolor}{rgb}{0.0,0.3,0.5}
\definecolor{dodgerblue}{HTML}{1E90FF}
\usepackage[unicode, colorlinks=true, linkcolor=linkcolor, citecolor=linkcolor, filecolor=linkcolor,urlcolor=linkcolor, pdfusetitle]{hyperref}
\usepackage[all]{hypcap}
\usepackage[T1]{fontenc}
\usepackage[utf8]{inputenc}
\usepackage{orcidlink}
\usepackage[normalem]{ulem}

\usepackage{graphicx}
\newcommand{\orcid}[1]{\href{https://orcid.org/#1}{\includegraphics[width=10pt]{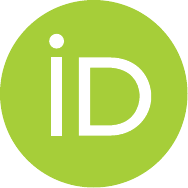}}}

\usepackage{lipsum}

\def\apjl{{Astrophys.~J.~Lett.}}
\def\apjs{{Astrophys.~J.~Supp.}}
\def\ao{{Appl.~Opt.}}

\def\aap{{Astron.~Astrophys.}}

\definecolor{romared}{RGB}{142,0,28}
\usepackage{hyperref}
\hypersetup{
    pdfnewwindow=true,      
    colorlinks=true,       
    linkcolor=romared,          
    citecolor=romared,        
    filecolor=romared,      
    urlcolor=romared        
}

\newcolumntype{Y}{>{\centering\arraybackslash}X}

\def\bi{\begin{itemize}[noitemsep,leftmargin=*]
\setlength\itemsep{1em}
        }
\def\ei{\end{itemize}}

\newcommand{\del}{\partial}
\renewcommand{\vec}[1]{\boldsymbol{#1}}
\newcommand{\beq}{\begin{eqnarray}}
\newcommand{\eeq}{\end{eqnarray}}

\makeatletter
\newcommand*{\balancecolsandclearpage}{\close@column@grid \cleardoublepage \twocolumngrid}
\makeatother

\begin{document}

\title{Subtracting Compact Binary Foregrounds to Search for Subdominant Gravitational-Wave Backgrounds in Next-Generation Ground-Based Observatories}

\author{Bei Zhou \orcid{0000-0003-1600-8835}}
\affiliation{William H. Miller III Department of Physics and Astronomy, Johns Hopkins University, Baltimore, Maryland 21218, USA}
\author{Luca Reali \orcid{0000-0002-8143-6767}}
\affiliation{William H. Miller III Department of Physics and Astronomy, Johns Hopkins University, Baltimore, Maryland 21218, USA}
\author{Emanuele Berti \orcid{0000-0003-0751-5130}}
\affiliation{William H. Miller III Department of Physics and Astronomy, Johns Hopkins University, Baltimore, Maryland 21218, USA}
\author{Mesut \c{C}al{\i}\c{s}kan \orcid{0000-0002-4906-2670}}
\affiliation{William H. Miller III Department of Physics and Astronomy, Johns Hopkins University, Baltimore, Maryland 21218, USA}
\author{Cyril Creque-Sarbinowski \orcid{0000-0002-6197-5421}}
\affiliation{William H. Miller III Department of Physics and Astronomy, Johns Hopkins University, Baltimore, Maryland 21218, USA}
\author{Marc Kamionkowski \orcid{0000-0001-7018-2055}}
\affiliation{William H. Miller III Department of Physics and Astronomy, Johns Hopkins University, Baltimore, Maryland 21218, USA}
\author{B. S. Sathyaprakash \orcid{0000-0003-3845-7586}}
\affiliation{Institute for Gravitation and the Cosmos, Department of Physics, Penn State University, University Park, Pennsylvania 16802, USA}
\affiliation{Department of Astronomy and Astrophysics, Penn State University, University Park, Pennsylvania 16802, USA}
\affiliation{School of Physics and Astronomy, Cardiff University, Cardiff, CF24 3AA, United Kingdom}

\date{\today}  

\begin{abstract}
Stochastic gravitational-wave backgrounds (SGWBs) derive from the superposition of numerous individually unresolved gravitational-wave (GW) signals. Detecting SGWBs provides us with invaluable information about astrophysics, cosmology, and fundamental physics.  In this paper, we study SGWBs from binary black-hole (BBH) and binary neutron-star (BNS) coalescences in a network of next-generation ground-based GW observatories (Cosmic Explorer and Einstein Telescope) and determine how well they can be measured; this then limits how well we can observe other subdominant astrophysical and cosmological SGWBs.  We simulate all-Universe populations of BBHs and BNSs and calculate the corresponding SGWBs, which consist of a superposition of (i) undetected signals, and (ii) the residual background from imperfect removal of resolved sources.  The sum of the two components sets the sensitivity for observing other SGWBs.  Our results show that, even with next-generation observatories, the residual background is large and limits the sensitivity to other SGWBs. The main contributions to the residual background arise from uncertainties in inferring the coalescence phase and luminosity distance of the detected signals.  Alternative approaches to signal subtraction would need to be explored to minimize the BBH and BNS foreground in order to observe SGWBs from other subdominant astrophysical and cosmological sources.
\end{abstract}

\maketitle

\section{Introduction}
\label{sec_intro}

The first detection of gravitational waves (GWs) from a binary black-hole (BBH) coalescence in 2015 has opened a new window to the Universe~\cite{LIGOScientific:2016aoc}. Soon after that, the first detection of GWs from a binary neutron-star (BNS) coalescence~\cite{LIGOScientific:2017vwq} and the observation of its electromagnetic counterpart~\cite{LIGOScientific:2017ync, LIGOScientific:2017zic} have significantly advanced the field of multimessenger astronomy.
So far, nearly 100 BBHs and 2 BNSs have been detected~\cite{LIGOScientific:2021djp,Nitz:2021zwj,Olsen:2022pin}.
These observations have made crucial contributions to astrophysics, cosmology, and fundamental physics~\cite{LIGOScientific:2021aug,
KAGRA:2021duu,
LIGOScientific:2021sio}. 

In addition to those loud and individually resolved GW events, a plethora of signals from multiple kinds of sources remain too weak to be detected, and their incoherent superposition gives rise to stochastic GW backgrounds (SGWBs)~\cite{Allen:1997ad,Sathyaprakash:2009xs,Caprini:2018mtu,Christensen:2018iqi,Renzini:2022alw}. 
A huge variety of SGWBs are expected. Some are of astrophysical origin, such as those from supernova explosions~\cite{Ferrari:1998ut, Buonanno:2004tp, Crocker:2015taa, Crocker:2017agi, Finkel:2021zgf} or the cumulative sum of unresolved compact-binary coalescences (CBCs)~\cite{Wu:2011ac,Marassi:2011si,Zhu:2011bd,Rosado:2011kv,Zhu:2012xw, Dominik:2014yma}. Others are of cosmological origin and include SGWBs from standard inflation~\cite{Grishchuk:1974ny,Starobinsky:1979ty,Grishchuk:1993te},  axion inflation~\cite{Barnaby:2011qe}, cosmic strings~\cite{Damour:2004kw, Siemens:2006yp, Olmez:2010bi, Regimbau:2011bm}, etcetera. The predicted energy densities of these different SGWBs vary by many orders of magnitude.

The detection and characterization of SGWBs has important scientific payoffs. Astrophysical backgrounds potentially contain key information about the mass and redshift distributions and other properties of their corresponding sources~\cite{LIGOScientific:2020kqk, KAGRA:2021duu, Bavera:2021wmw}. 
In addition, observing cosmological SGWBs would open up a unique window to the earliest moments of the Universe and to the physical laws that apply at the highest energy (up to the limits of the Planck scale)~\cite{Grishchuk:1974ny,Starobinsky:1979ty,Grishchuk:1993te, Barnaby:2011qe, Damour:2004kw, Siemens:2006yp, Olmez:2010bi, Regimbau:2011bm}.
The current second-generation detector network (i.e., advanced LIGO~\cite{Harry:2010zz, LIGOScientific:2014pky}, advanced Virgo~\cite{VIRGO:2014yos}, and KAGRA~\cite{Aso:2013eba}) did not detect any SGWB in their searches~\cite{KAGRA:2021kbb}, putting upper bounds on the amplitude of the energy spectrum in various frequency bands~\cite{KAGRA:2021kbb,KAGRA:2021duu}. The SGWB for current detectors is projected to be dominated by the CBC background, as their network sensitivity allows only a small fraction of them to be resolved and subtracted. Therefore, detection of the subdominant astrophysical and cosmological SGWBs cannot be accomplished.

The situation could be significantly improved in next-generation (XG) observatories, including Cosmic Explorer (CE)~\cite{Reitze:2019iox} and the Einstein Telescope (ET)~\cite{Punturo:2010zz}. They are expected to detect hundreds of thousands of BBHs and BNSs per year at a signal-to-noise ratio (SNR) larger than 12 (this is the standard SNR threshold used by LIGO and Virgo, although the optimal choice may be different for XG detectors~\cite{Borhanian:2022czq,Ronchini:2022gwk,Iacovelli:2022bbs}). Meanwhile, the individually resolvable sources will be much better measured, and thus more precisely subtracted as a foreground. 

In this work, we study the CBC background in XG ground-based GW observatories. We consider populations of BNSs and BBHs with local merger rates consistent with the latest LIGO/Virgo/KAGRA (LVK) catalog~\cite{LIGOScientific:2021djp,KAGRA:2021duu}. We study how well the individually resolvable events can be measured and subtracted, and how much the unavoidable contribution from imperfect subtraction ($\Omega_{\rm err}$) contributes to the remaining SGWBs of CBCs. The other contribution is from the superposition of unresolved CBC GW signals ($\Omega_{\rm unres}$). The sum of these two residual backgrounds sets the effectivity sensitivity and determines how well other subdominant SGWBs can be detected.  
It is thus essential to understand these backgrounds and to think about how they can be minimized. In this paper we estimate the optimal signal-to-noise ratio threshold at which the sum $\Omega_{\rm unres} + \Omega_{\rm err}$ is minimized for BBH and BNS populations, respectively.

To our knowledge, the problem of subtracting the CBC foreground from resolved sources to detect subdominant SGWBs was first studied in detail in the context of the Big Bang Observer (BBO), a space-based interferometer network concept to observe primordial GWs~\cite{Cutler:2005qq, Cutler:2009qv}. After the detection of GW150914, various authors considered the possibility that BBHs may contribute to the confusion noise for the space-based detector LISA~\cite{Nishizawa:2016jji} and possible methods for their subtraction~\cite{Pieroni:2020rob}. More recent studies have applied an information-matrix formalism to the case of XG observatories~\cite{Regimbau:2016ike, Sachdev:2020bkk}.  Here (and in the companion paper~\cite{Zhou:2022otw}) we build upon the prior work of Refs.~\cite{Regimbau:2016ike, Sachdev:2020bkk} by expanding the range of binary parameters assumed to be determined for each GW signal. While those studies focused on the effect of the three dominant phase parameters (the redshifted detector-frame chirp mass $\mathcal{M}_z$, coalescence phase $\phi_c$, and time of coalescence $t_c$), we consider the larger 9-dimensional parameter space characterizing nonspinning binaries. We find that correlations and degeneracies between different parameters, and in particular the uncertainty in the amplitude of the individual signals, result in a much larger value for $\Omega_{\rm err}$.

The rest of the paper is organized as follows. In Sec.~\ref{sec_pop}, we present the assumptions on the mass and redshift distributions and local merger rates that underlie our simulation of BBH and BNS coalescence events.  In Sec.~\ref{sec_GWcalc}, we detail the formalism and assumptions (on waveform models, detector networks, and parameter estimation) that we use to compute SGWBs, and in particular $\Omega_{\rm unres}$ and $\Omega_{\rm err}$.  In Sec.~\ref{sec_rslt} we present our results, and in Sec.~\ref{sec_concl} we discuss their implications and possible directions for future work.

Throughout this paper $G$ is the gravitational constant, $c$ the speed of light, $H_0$ the Hubble constant, and we use the $\Lambda$CDM cosmological model with cosmological parameters taken from Planck 2018~\cite{Planck:2018vyg}.

\section{Simulating the Compact Binary Population}
\label{sec_pop}

In general, the GW signal emitted by a non-eccentric BBH (or BNS with tidal interactions neglected) can be characterized by a set of $15$ parameters. 
\begin{itemize}
    \item Eight intrinsic parameters, which includes the two independent mass parameters and two spin vectors ($\vec{\chi}_{1,2}$, where 1 stands for the primary while 2 for the secondary). For the masses, one uses either the companion masses $\{m_1, m_2\}$ or $\{\mathcal{M}, \eta \}$, where $\mathcal{M}$ is the chirp mass and $\eta$ is the symmetric mass ratio. We use the former combination when generating the astrophysical populations, and the latter when calculating the information matrix.
    
    \item Seven extrinsic parameters, which includes the merger redshift $z$ (or luminosity distance $D_L$), the sky position of the source (right ascension $\alpha$ and declination $\delta$), the inclination angle $\iota$, the polarization angle $\psi$, the coalescence phase $\phi_c$, and the coalescence time $t_c$.
\end{itemize}

\begin{table}[t!]
    \centering
    \begin{tabular}{c|c|c} \hline \hline
        Parameter &  BNS &  BBH   \\ \hline
        $m_1$    &  Double Gaussian \cite{Farrow:2019xnc} & \multirow{2}{*}{\texttt{POWER+PEAK}  \cite{KAGRA:2021duu}} \\ 
        $m_2$   & Uniform $[1.14,1.46]~M_{\odot}$ \\ \hline
        $\vec{\chi}_{1}$, $\vec{\chi}_{2}$ & \multicolumn{2}{c}{$\vec{0}$}  \\ \hline
        \multirow{2}{*}{$z$} & \multirow{2}{*}{SFR~\cite{Vangioni:2014axa} \texttt{+} time delay} & \multirow{2}{*}{\shortstack[c]{SFR~\cite{Vangioni:2014axa} \texttt{+} time delay \\ \texttt{+} metallicity}} \\ & & \\ \hline
        $\cos \iota$ & \multicolumn{2}{c}{\multirow{2}{*}{Uniform in $[-1,1]$}} \\ $\cos \delta$ \\ \hline
        $\alpha$, $\psi$, $\phi_c$ &
        \multicolumn{2}{c}{Uniform in $[0,2\pi]$}
        \\
        \hline
        $t_c$ & \multicolumn{2}{c}{0} \\ \hline \hline
    \end{tabular}
    \caption{Distributions for the parameters of our BNS and BBH populations. See text for the details. 
    } 
    \label{tab_pop}
\end{table}

Table~\ref{tab_pop} summarizes the distributions of the above parameters for our BBH and BNS populations. We specify the parameters for the mass and redshift distributions in Secs.~\ref{sec_pop_mass} and \ref{sec_pop_redshift}, respectively. The effect of spins is expected to be subdominant, so we set the spins to zero for all binaries. Moreover, we use uniform distributions for $\cos {\iota}$, $\cos \delta$, $\alpha$, $\psi$, and $\phi_c$. We use zero as the initial (true) value of the coalescence time $t_c$ for all the binaries, but we include uncertainties on its recovered value from parameter estimation. The choice of $t_c=0$ as the initial value does not induce a biased result, because sampling the sources isotropically in the sky is effectively degenerate with assuming a uniform distribution in $t_c$. In Sec.~\ref{sec_pop_rate} we discuss the local merger rates of BBHs and BNSs, which sets the normalization of the redshift distribution (Sec.~\ref{sec_pop_redshift}).


\subsection{Mass distribution}
\label{sec_pop_mass}

We model the masses of our BBH population using the \texttt{POWER+PEAK} phenomenological model described by the LVK Collaboration~\cite{KAGRA:2021duu}, which performed a population study based on the third Gravitational-Wave Transient Catalog (GWTC-3).
This model gives the highest Bayes factor among all the options considered there.\footnote{We note that the population parameters of this model have been slightly revised in the latest versions of the preprint~\cite{KAGRA:2021duu}. However, to this day the public data release~\cite{ligo_scientific_collaboration_and_virgo_2021_5655785} has not been updated since the first version, and thus we adopt those parameters for our study~\cite{Weinstein}. The impact on our analysis of the slight change in the population parameters is negligible.}
The primary black-hole (BH) mass follows a truncated power law with the addition of a Gaussian peak and an exponential tapering at low masses, i.e., 
\begin{align}
P(m_1) & \propto [ (1-\lambda)P_{\rm law}(m_1|\gamma_1,m_{\rm max}) \notag \\
 & + \lambda G(m_1|\mu_m,\sigma_m) ] \, S(m_1|m_{{\rm min}}, \delta_m)\,,
\end{align}
where $P_{\rm law}(m_1|\gamma_1,m_{\rm max})$ is a power-law distribution with slope $\gamma_1=-3.40$ and cutoff at $m_{\rm max}=86.85\, M_\odot$, $G(m_1|\mu_m,\sigma_m)$ a Gaussian distribution with mean $\mu_m=33.73\, M_\odot$ and standard deviation $\sigma_m=3.36\, M_\odot$, and $S(m_1|m_{\rm min},\delta_m)$ is a tapering function which rises monotonically from $0$ to $1$ within $[m_{\rm min},m_{\rm min}+\delta_m]$, where $m_{\rm min}=5.08\, M_\odot$ and $\delta_m=4.83\, M_\odot$ (see Appendix B of Ref.~\cite{KAGRA:2021duu} for details). 

The secondary BH mass is derived from the mass ratio $q=m_2/m_1$, which is sampled, for each binary, according to the smoothed power law
\beq
P(q) \propto q^{\gamma_q} S(m_1 q|m_{\rm min}, \delta_m) \,,
\eeq
with $\gamma_q=1.08$.  

The masses of our BNS population are distributed according to the preferred model from Ref.~\cite{Farrow:2019xnc}. They performed a Bayesian analysis on a sample of $17$ Galactic BNSs and showed that the primary (recycled) neutron-star (NS) mass follows a double Gaussian distribution,
\begin{align}
P(m_1) = &\frac{\gamma_{\rm NS}}{\sqrt{2 \pi}\sigma_1}
e^{\tfrac{(m_1-\mu_1)^2}{2\sigma_1^2}}       
+ \frac{1-\gamma_{\rm NS}}{ \sqrt{2 \pi}\sigma_2} e^{\tfrac{(m_1-\mu_2)^2}{2\sigma_2^2}}\,,
  \label{eqn:DG}
\end{align}
with $\mu_1=1.34\,M_{\odot}$, $\mu_2=1.47\,M_{\odot}$, $\sigma_1=0.02\,M_{\odot}$, $\sigma_2=0.15\,M_{\odot}$, and $\gamma_{\rm NS}=0.68$. The secondary (non-recycled) NS mass is instead distributed uniformly within the range $m_2\in[1.14,1.46]M_{\odot}$. 


\subsection{Redshift distribution}
\label{sec_pop_redshift}
We simulate BBHs and BNSs up to redshift $z=10$. The redshift distribution we use comes from Refs.~\cite{Regimbau:2016ike, Sachdev:2020bkk}, as detailed below. 

First of all, we assume the binary formation rate to follow the star-formation rate (SFR)~\cite{Vangioni:2014axa},
\begin{equation}
R_{\rm sf}(z_f)=\nu \frac{a \, e^{b\left(z_f-z_p\right)}}{a-b+b \, e^{\left[ a\left(z_f-z_p\right)\right]}} \,,
\end{equation}
where $\nu = 0.146\, M_\odot \rm \, yr^{-1}\, Mpc^3$, $z_p = 1.72$, $a = 2.80$, and $b = 2.46$. The functional form is taken from Ref.~\cite{Springel:2002ux}, with the best-fit parameters from high-redshift star-formation data based on gamma-ray bursts, and normalization from Refs.~\cite{Trenti:2013oaa, Behroozi:2014tna}.

In addition, we set a metallicity cut on BBHs, as massive BHs are more likely to have formed in low-metallicity environments~\cite{LIGOScientific:2016fpe, LIGOScientific:2016vpg}. For binaries with at least one BH heavier than $30\, M_\odot$, we reweigh the star-formation rate by the fraction of stars with metallicities less than half of the solar metallicity~\cite{LIGOScientific:2017zlf}. The metallicities of stars are drawn from a $\log_{10}$-normal distribution with standard deviation of 0.5 and redshift-dependent mean value from Ref.~\cite{Madau:2014bja}, rescaled upwards by a factor of 3 to account for local observations~\cite{Vangioni:2014axa, Belczynski:2016obo}.

The time delay $t_d$ between the formation of a binary and its merger has probability distribution $p(t_d)$. We assume $p(t_d) \propto 1/t_d$ in the range $t_d^{\min} < t_d < t_d^{\max}$.  We set $t_d^{\max}$ to be the Hubble time~\cite{Belczynski:2001uc, Ando:2004pc, Belczynski:2006br, Berger:2006ik, Nakar:2007yr, OShaughnessy:2007brt, Dominik:2012kk, Dominik:2013tma}, while we assume $t_d^{\min}=50$\,Myr for BBHs~\cite{Dominik:2013tma, LIGOScientific:2016fpe, LIGOScientific:2017zlf}, and $t_d^{\min}=20$\,Myr for BNSs~\cite{Meacher:2015iua, LIGOScientific:2017zlf}.

Combining all of the assumptions above leads to the merger rate per comoving volume in the source frame
\begin{equation}
R_m(z)=\int_{t_d^{\min}}^{t_d^{\max}} R_{\rm sf}\{\tilde{z}[\tilde{t}(z)-t_{d}]\} \, p(t_{d}) \, d t_{d} \,,
\end{equation}
where $z$ is the merger redshift, $\tilde{t}(z)$ is the cosmic time when the merger happens,  $\tilde{z}$ is the redshift as a function of cosmic time, and $\tilde{t}(z)-t_{d}$ is the cosmic time when the binary forms. 

Finally, the merger rate in the observer frame is
\begin{equation}
R_z(z)=\frac{R_m(z)}{1+z} \frac{d V_c}{d z}(z) \, ,
\label{eq_Rz}
\end{equation}
where $dV_c/dz$ is the comoving volume element.  

\subsection{Local rates and astrophysical uncertainty}
\label{sec_pop_rate}

The normalization of $R_z(z)$ in Eq.~\eqref{eq_Rz}, i.e., $R_m(z=0)$, is set by the measured local merger rate from LVK observations~\cite{KAGRA:2021duu,LIGOScientific:2020kqk}. For each of our binary populations, we choose a fiducial value and a $90\%$ confidence interval based on the GWTC-3 catalog~\cite{KAGRA:2021duu} to characterize the uncertainties.

For BBHs, the fiducial value we use is the best estimate for the \texttt{POWER+PEAK} model from GWTC-3~\cite{KAGRA:2021duu}, i.e., $R_m(z=0)=28.3~\rm Gpc^{-3}yr^{-1}$. The $90\%$ confidence interval is $[17.9,44]\, \rm Gpc^{-3}yr^{-1}$, covering several different astrophysical models~\cite{KAGRA:2021duu}. 

For BNSs, we use a more sophisticated mass distribution that is different from those used in Ref.~\cite{KAGRA:2021duu}. However, Ref.~\cite{KAGRA:2021duu} considers several different models, so their combined 90\% confidence interval, $[10,1700]\,\rm Gpc^{-3}yr^{-1}$, should mostly cover that of our model. The large uncertainty is mainly because there are only two BNS coalescence events detected so far~\cite{LIGOScientific:2017vwq,LIGOScientific:2020aai}. 
For the fiducial value, we use the best estimate from GWTC-2~\cite{LIGOScientific:2020kqk}, $320~\rm Gpc^{-3}yr^{-1}$. This number is well within the uncertainty range quoted above and closer to the estimates considered in recent forecasts for XG detectors~\cite{Borhanian:2022czq,Ronchini:2022gwk,Iacovelli:2022bbs}, allowing for more direct comparisons with the literature.

\section{Calculation of the Stochastic Gravitational-Wave Background from Compact Binaries}
\label{sec_GWcalc}

The energy-density spectrum of a SGWB can be described by the dimensionless quantity~\cite{Allen:1997ad}
\begin{equation}
\Omega_{\rm GW}(f) \equiv
\frac{f}{\rho_c} \frac{d\rho_{\rm GW}}{d f}(f) 
=  
\frac{1}{\rho_c c} f F(f) 
\,,
\label{eq_omegadef}
\end{equation}
where $f$ is the GW frequency, $\rho_{\rm GW}$ the GW energy density, $F$ the GW energy flux, and $\rho_c=3H_0^2/8\pi G$ is the critical density of the Universe. 

In this section, we summarize the formalism we used to compute the CBC SGWBs. We first give an overview of the SGWB in Sec.~\ref{sec_GWcalc_SGWB}, including its different components. Then, we present the detection signal-to-noise ratio (SNR) of a GW signal in Sec.~\ref{sec_GWcalc_SNR}, and the statistical uncertainty due to imperfect subtraction of the resolved signals in Sec.~\ref{sec_GWcalc_err}. Finally, in Sec.~\ref{sec_GWcalc_others} we discuss the systematic effects due to waveform modeling uncertainties and the XG detector networks considered in the analysis. For some of our calculations, including the waveforms, the detection SNR, and the information matrix of each CBC event, we use GWBENCH~\cite{Borhanian:2020ypi}, 
a Python package designed for GW information matrix analyses given a network of GW detectors.

\subsection{Contributions to the stochastic gravitational-wave background}
\label{sec_GWcalc_SGWB}
The total energy flux from a given population of sources (BBHs or BNSs) is given by~\cite{Allen:1997ad}
\begin{equation}
F_{\rm tot}(f)=T^{-1} \frac{\pi c^{3}}{2 G} f^{2} \sum_{i=1}^{N}\left[|\tilde{h}_{+}^i(f)|^{2}+|\tilde{h}_{\times}^i(f)|^{2}\right] \,,
\label{eq_flux}
\end{equation}
where the index $i$ runs over the $N$ sources, and $\tilde{h}_{+}^i$ and $\tilde{h}_{\times}^i$ are the two polarization modes ($+$ and $\times$) of the waveform in the Fourier domain. The number of sources $N$ is proportional to the total observation time $T$.\footnote{To make sure that our computed backgrounds are convergent we consider BBH and BNS samples with $N=10^5$ events in each case, corresponding to $T \simeq 2.6$~years (for BBHs) and $0.2$~years (for BNSs) of observations.}

The energy-density spectrum $\Omega_{\rm tot}(f)$ of those sources can then be calculated using Eq.~\eqref{eq_omegadef}. This quantity encompasses the contribution from all the sources in a population. The sources are individually resolvable if their detection SNR, defined in Sec.~\ref{sec_GWcalc_SNR} below, is larger than a threshold value $\rm SNR_{thr}$.
The contribution of resolvable sources to the total SGWB, $\Omega_{\rm res}$, can in principle be subtracted, leaving an incoherent sum of unresolved signals, $\Omega_{\rm unres}$, that contributes to the SGWB. The unresolved contribution $\Omega_{\rm unres}$ can be calculated in the same way as $\Omega_{\rm tot}$ [Eqs.~\eqref{eq_omegadef} and \eqref{eq_flux}], replacing the sum over all sources with a sum over individually unresolvable sources.

However, the subtraction of the resolvable component $\Omega_{\rm res}$ is always imperfect: the error in the characterization of the sources produces an additional contribution to the background, $\Omega_{\rm err}$. The reason is that the parameters of a resolvable event are never recovered perfectly due to detector noise. In other words, when we fit for the resolved signals and remove them from the data to get the remaining background, each of them leaves behind an unfitted residual in the data. Thus $\Omega_{\rm err}$ also contributes, effectively, to the unresolved SGWB (see Sec.~\ref{sec_GWcalc_err} for details). 
In summary, the total CBC SGWB residual foreground is given by $\Omega_{\rm err} + \Omega_{\rm unres}$.

Future XG observatories can resolve a very large number of individual sources, mostly BBHs and BNSs. The goal of the analysis presented below is to understand what types of SGWBs (other than those produced by CBCs) could be detected after subtraction. The sensitivity to these ``other'' SGWBs is set by the sum $\Omega_{\rm err} + \Omega_{\rm unres}$, which should be \textit{minimized}.

\subsection{Detection signal-to-noise ratio}
\label{sec_GWcalc_SNR}

The detectability of a GW signal can be assessed by computing its matched-filtered SNR. For a single GW detector, the SNR is defined by
\begin{equation}
    {\rm SNR}^2 = 4\int_{0}^{\infty}\frac{|\tilde{h}(f)|^2}{S_{n}(f)}\,df \,, 
    \label{eq_snr}
\end{equation}
where $S_n(f)$ is the power spectral density (PSD) of the detector (Sec.~\ref{sec_GWcalc_others}). 
For a network of $N_{\rm det}$ GW detectors, assuming uncorrelated noise between different detectors, the matched-filtered SNR of a GW signal is given by summing the individual-detector SNRs in quadrature:
\begin{equation}
    {\rm SNR} = \sqrt{\sum_{j=1}^{N_{\rm det}}{\rm SNR}_j^2} \,,
    \label{eq_snrnet}
\end{equation}
where $j$ labels each of the $N_{\rm det}$ detectors in the network.

We consider a signal resolved if its network SNR is larger than a certain threshold ${\rm SNR}_{\rm thr}$. For the LIGO/Virgo network the typical threshold for detection is ${\rm SNR}_{\rm thr}=12$, while for a network of XG detectors it is still under discussion. For this reason we will consider different choices for ${\rm SNR}_{\rm thr}$. In fact, one of the main goals of our analysis is to find the optimal value of ${\rm SNR}_{\rm thr}$ that allows us to best subtract the compact binary foreground.

\subsection{$\Omega_{\rm err}$ from imperfect subtraction}
\label{sec_GWcalc_err}

As mentioned above, the imperfect subtraction of individually resolved GW signals leaves behind a residual, $\Omega_{\rm err}$, that contributes to the total SGWB. This can be calculated as
\begin{equation}
\Omega_{\text{error}} = \frac{1}{\rho_{c}c} f F_{\text{error}}(f)\,,
\label{eq_Omega_err}
\end{equation}
with
\beq
\begin{aligned}
F_{\rm error}(f) &= T^{-1} \frac{\pi c^{3}}{2 G} f^2 \sum_{i=1}^{N_{\rm res}} \left[ \left| \tilde{h}_+(\vec{\theta}^i_{\rm tr};f) - \tilde{h}_+(\vec{\theta}^i_{\rm rec};f) \right|^2 \right. \notag \\ 
&+ \left. \left| \tilde{h}_\times(\vec{\theta}^i_{\rm tr};f) - \tilde{h}_\times(\vec{\theta}^i_{\rm rec};f) \right|^2 \right] \,,
\label{eq:Ferr}
\end{aligned}
\eeq
where the vector $\vec{\theta}^i$ represents the parameters of source $i$, while $\vec{\theta}^i_{\rm tr}$ and $\vec{\theta}^i_{\rm rec}$ are the true and recovered parameters, respectively.
The sum is computed only over the $N_{\rm res}< N$ resolved sources such that ${\rm SNR}_i > \rm SNR_{th}$. 

Ideally, one could do a full Bayesian parameter estimation on each resolved source, from which the recovered parameters are distributed according to a certain posterior probability. However, this method is impractical because of the size of our sample (we typically have $N\sim 10^5$).

We thus provide an estimate of the errors by adopting the linear signal approximation~\cite{Finn:1992wt}. We assume the posterior probability distribution for each source to be a multivariate Gaussian centered at the true parameters $\vec{\theta}^i_{\rm tr}$ with covariance matrix $\Sigma = \Gamma^{-1}$, where $\Gamma$ is the information matrix.
For a single GW detector, $\Gamma$ is defined as

\begin{equation}
    \Gamma_{\alpha\beta} = \left(\frac{\del h}{\del\theta^\alpha} \middle| \frac{\del h}{\del\theta^\beta}\right) \,,
\end{equation}
with $(\cdot|\cdot)$ the usual signal inner product,
\begin{equation}
    (a|b) = 4\,\mathrm{Re} \int_0^{\infty}\frac{\tilde{a}(f)\tilde{b}^*(f)}{S_n(f)}\,df \,,
\end{equation}
where $a$ and $b$ are two generic signals and we denote the complex conjugate with an asterisk. For a network of $N_{\rm det}$ detectors with uncorrelated noise, the total information matrix is given by the sum of single-detector information matrices
\begin{equation}
    \Gamma = \sum_{j=1}^{N_{\rm det}} \Gamma_j \,.
    \label{eq_infonet}
\end{equation}
%
\begin{figure}[t!]
\includegraphics[width=\columnwidth]{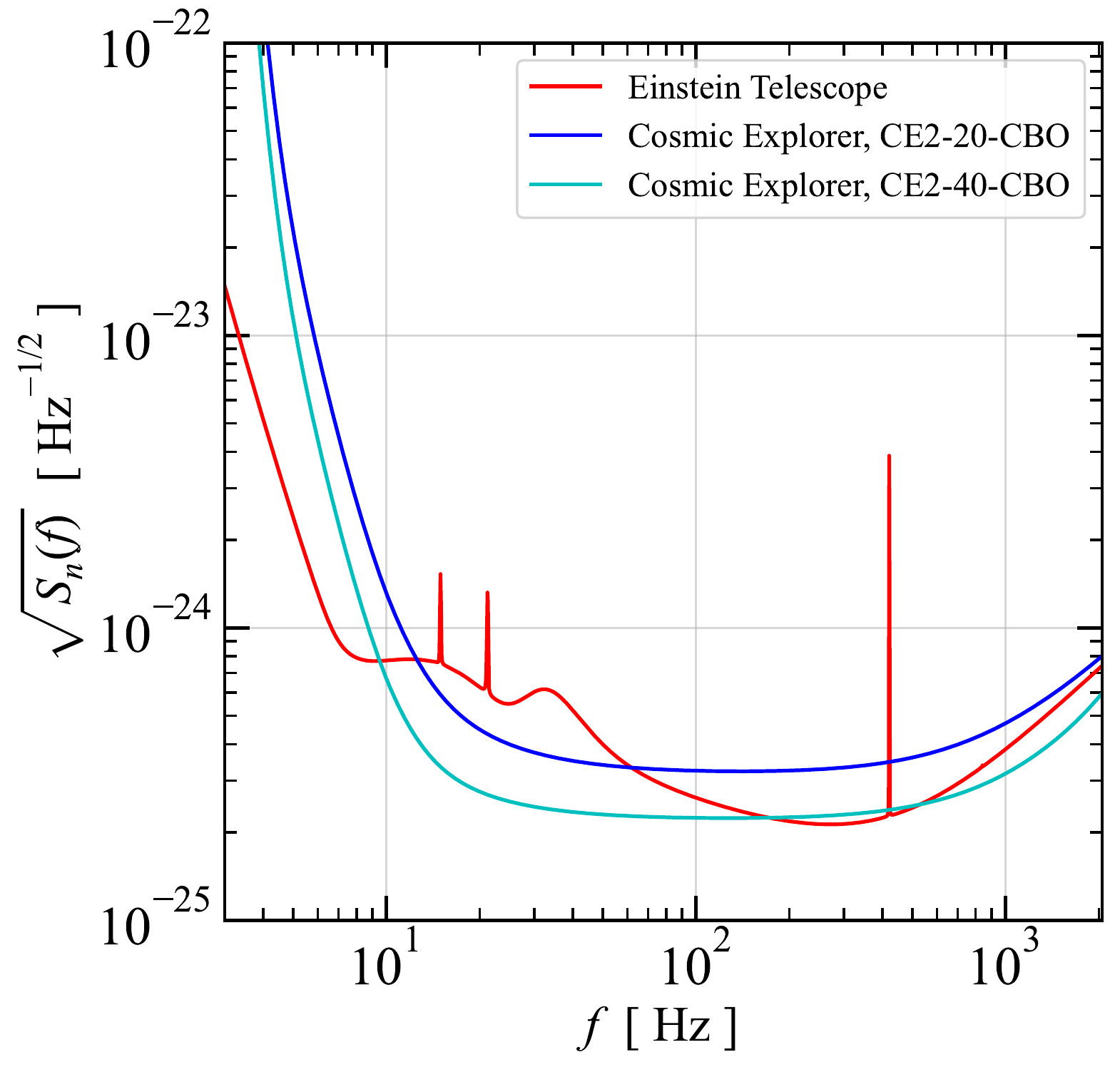}
\caption{Power spectral densities $\sqrt{S_n(f)}$ for the detectors in our network. The curve for ET represents the sensitivity of the whole triangular configuration~\cite{Punturo:2010zz,Borhanian:2020ypi}.
}
\label{fig_psds}
\end{figure}
\begin{figure*}[t!]
\includegraphics[width=\columnwidth]{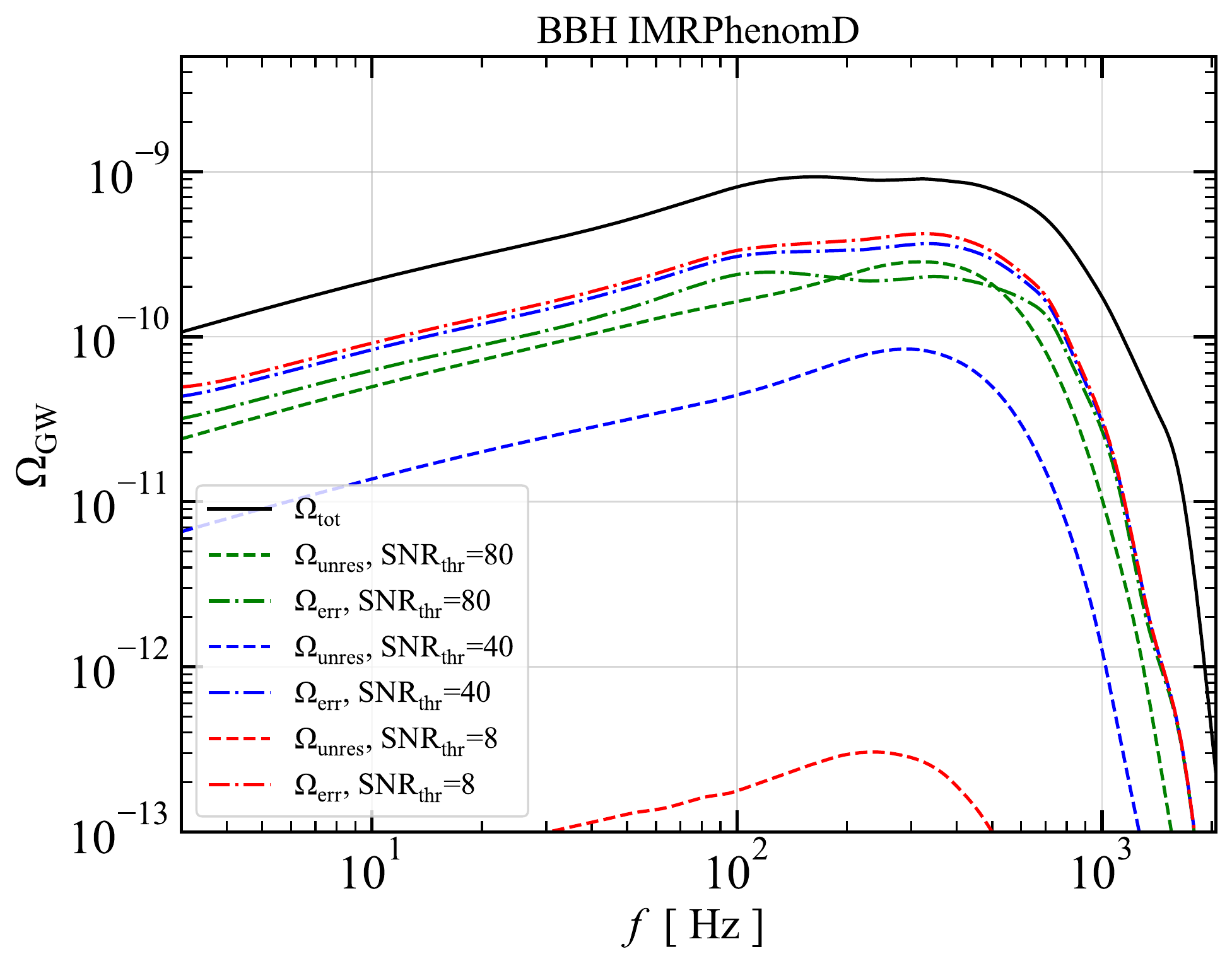}
\includegraphics[width=\columnwidth]{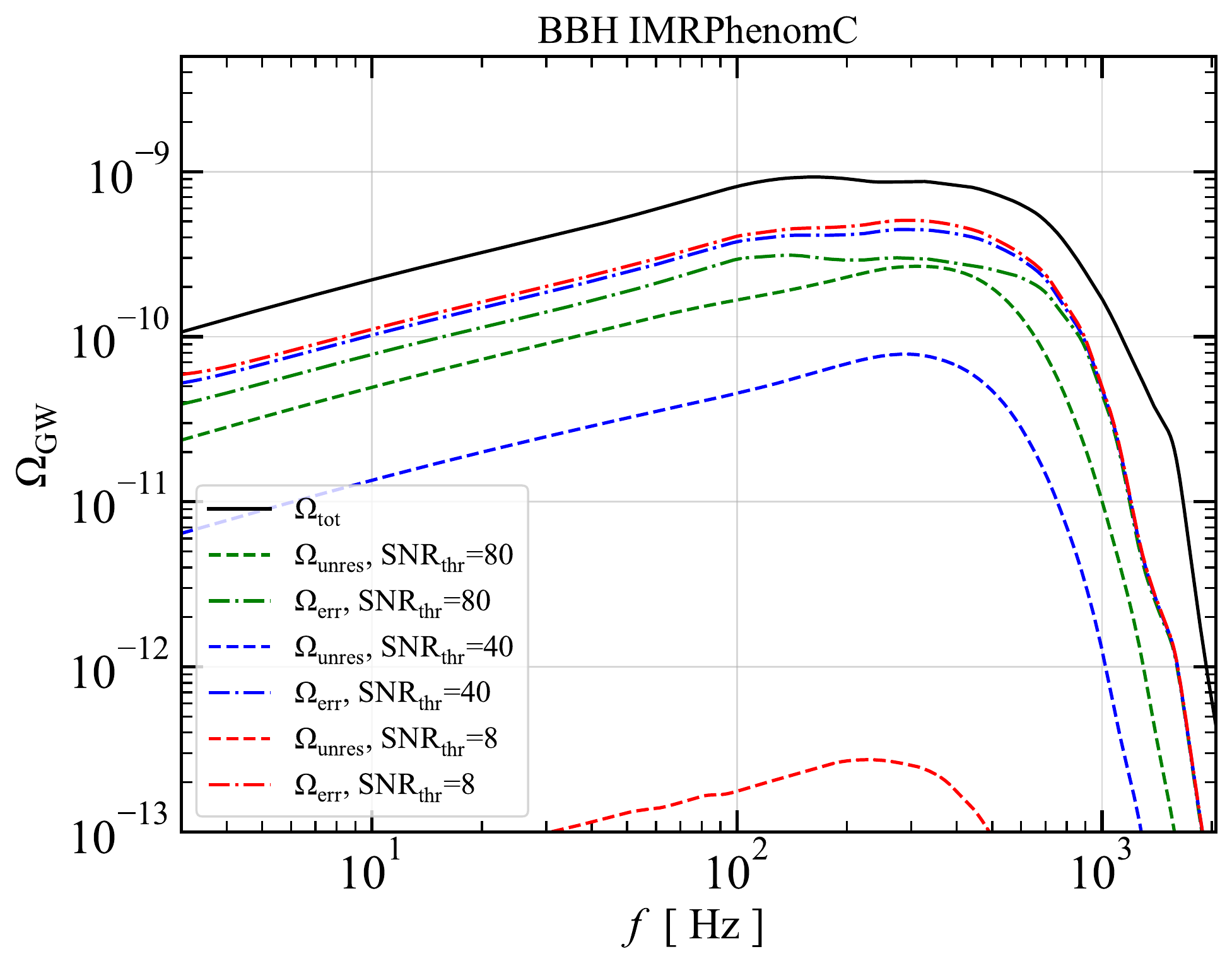}
\includegraphics[width=\columnwidth]{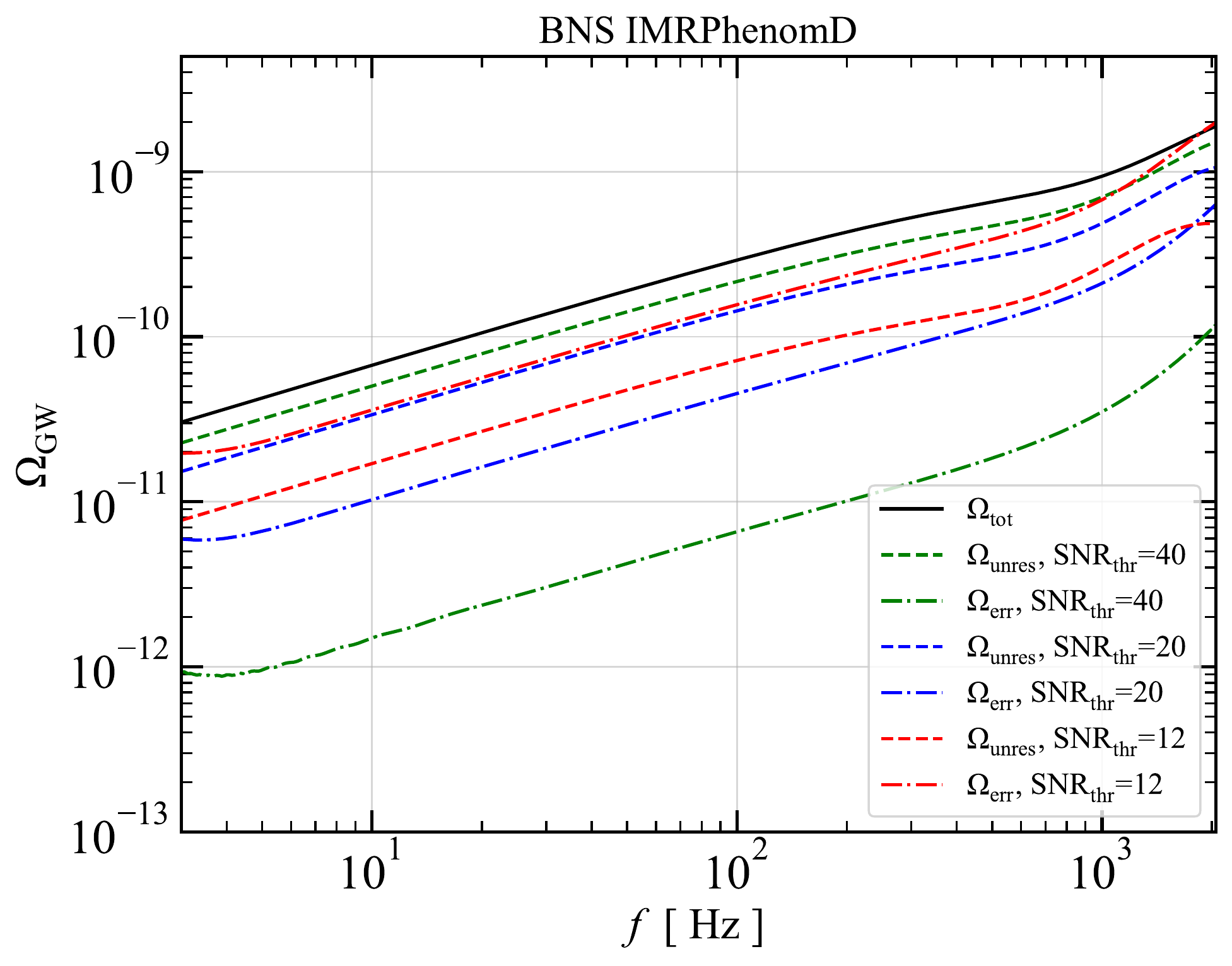}
\includegraphics[width=\columnwidth]{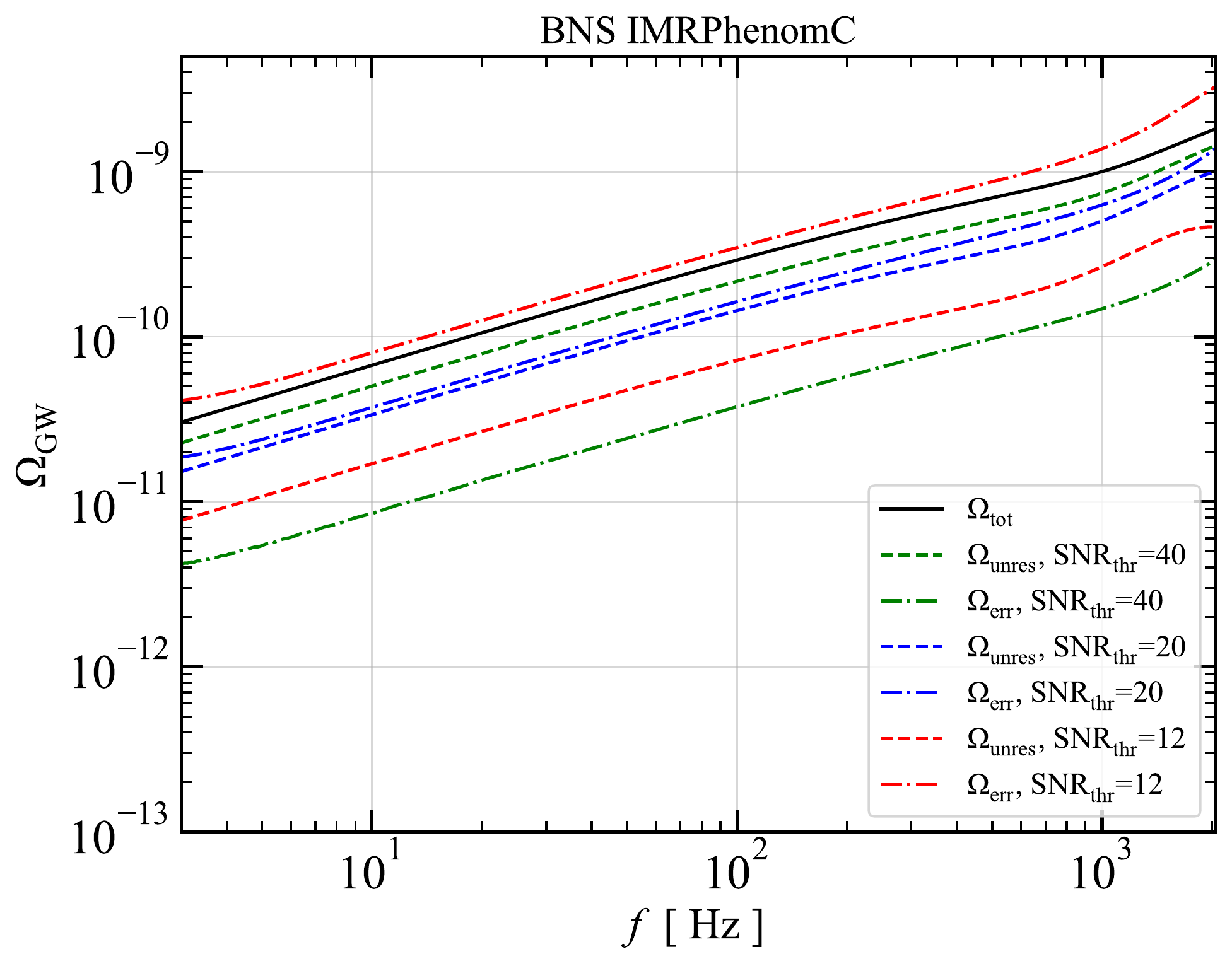}

\caption{Energy-density spectra $\Omega_{\rm tot}$ (black solid lines), $\Omega_{\rm unres}$ (dashed lines), and $\Omega_{\rm err}$ (dot-dashed lines) assuming our fiducial 3-detector network and different SNR thresholds, as indicated in the legend. We show results for BBHs (top panels) and BNSs (bottom panels) using two different waveform models, {\tt IMRPhenomD} (left panels) and {\tt IMRPhenomC} (right panels).}
\label{fig_Omegas}
\end{figure*}

\begin{figure}[h!]
\includegraphics[width=\columnwidth]{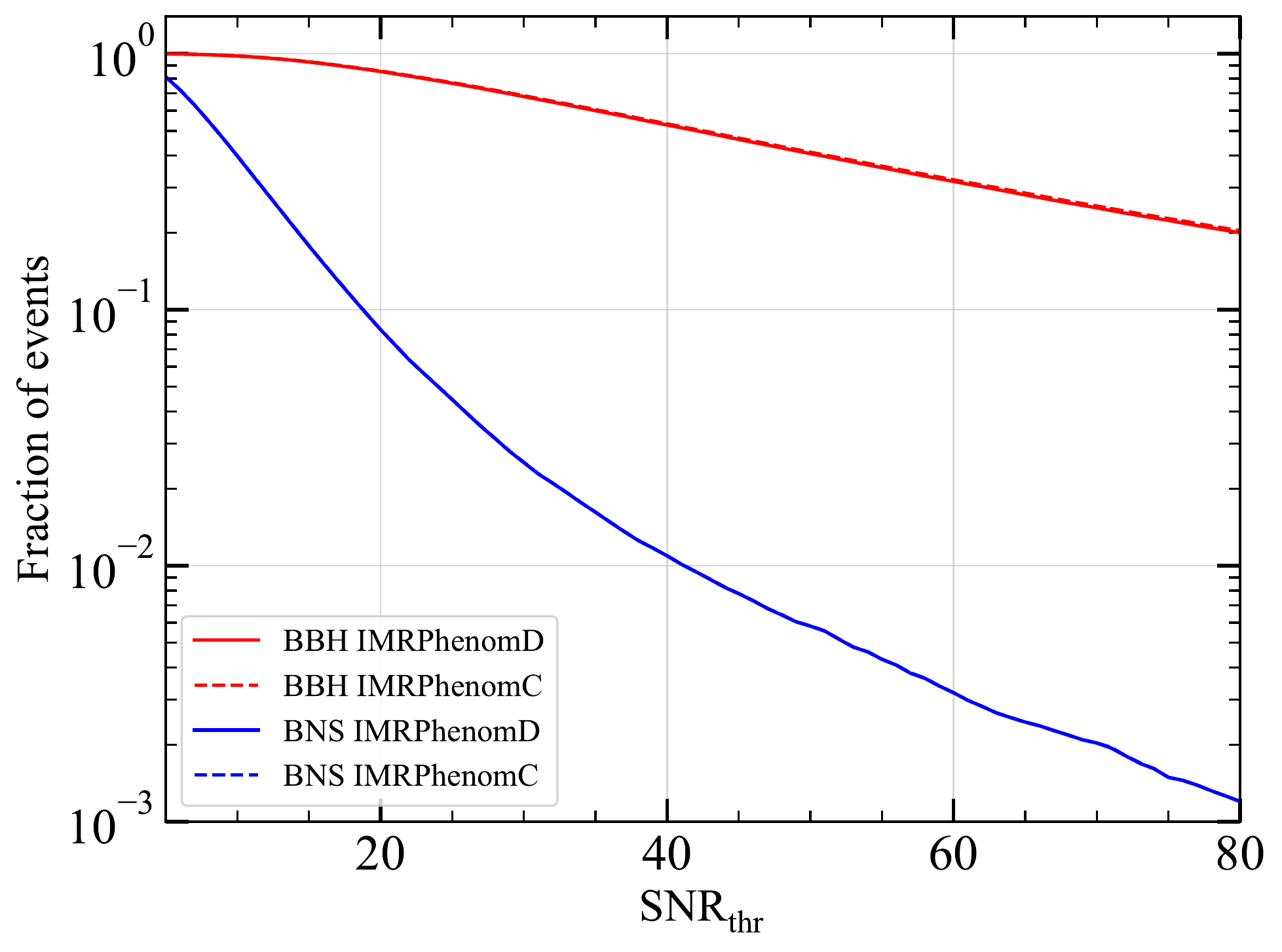}
\caption{Fraction of resolved events for BBHs (in red) and BNSs (in blue) as a function of $\rm SNR_{thr}$ for our fiducial 3-detector network. Solid (dashed) lines refer to {\tt IMRPhenomD} ({\tt IMRPhenomC}), and they show that the difference between waveform models is negligible from the point of view of detection.}
\label{fig_fraction_resolved}
\end{figure}

\begin{figure*}
\includegraphics[width=\columnwidth]{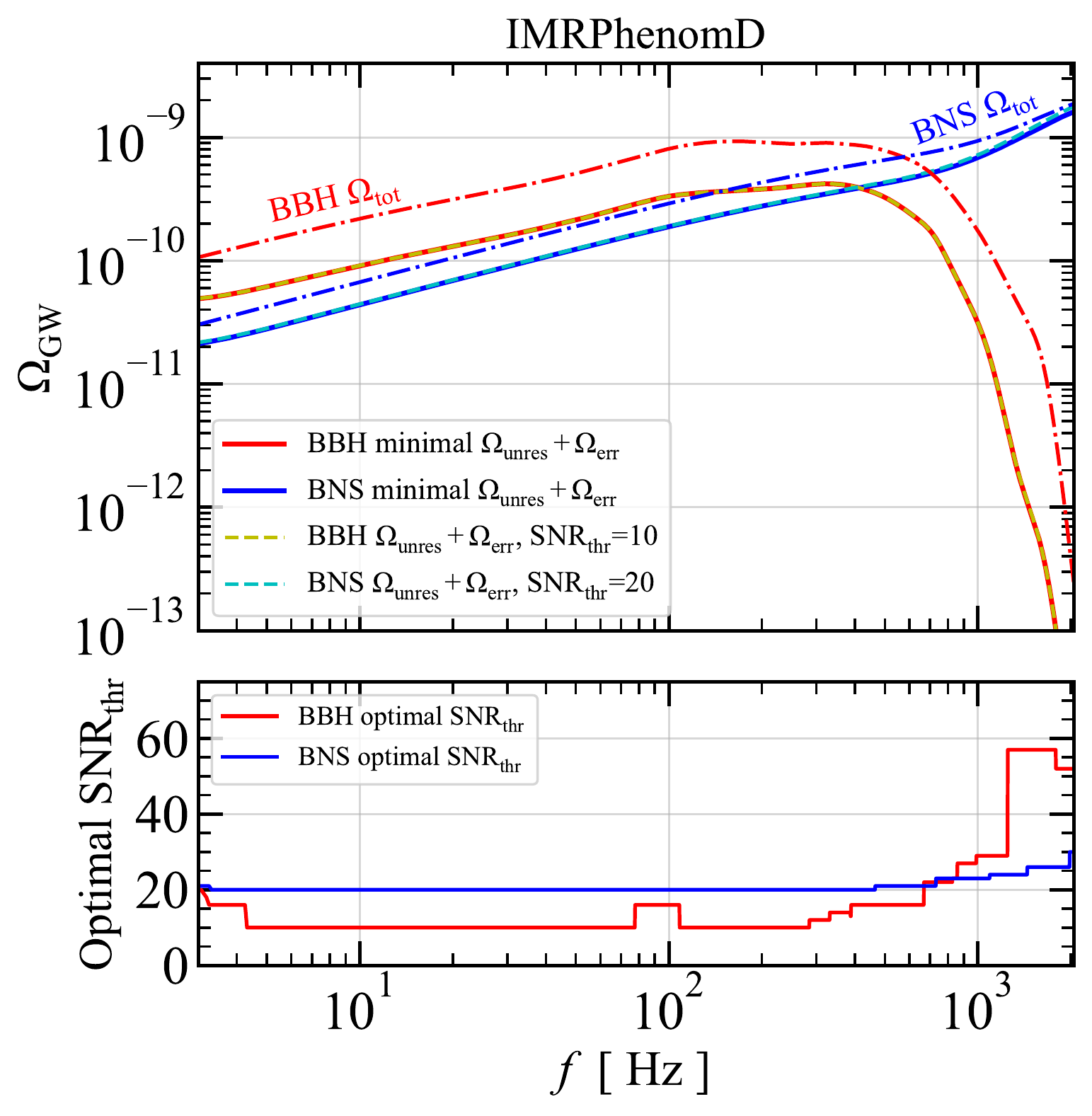}
\includegraphics[width=\columnwidth]{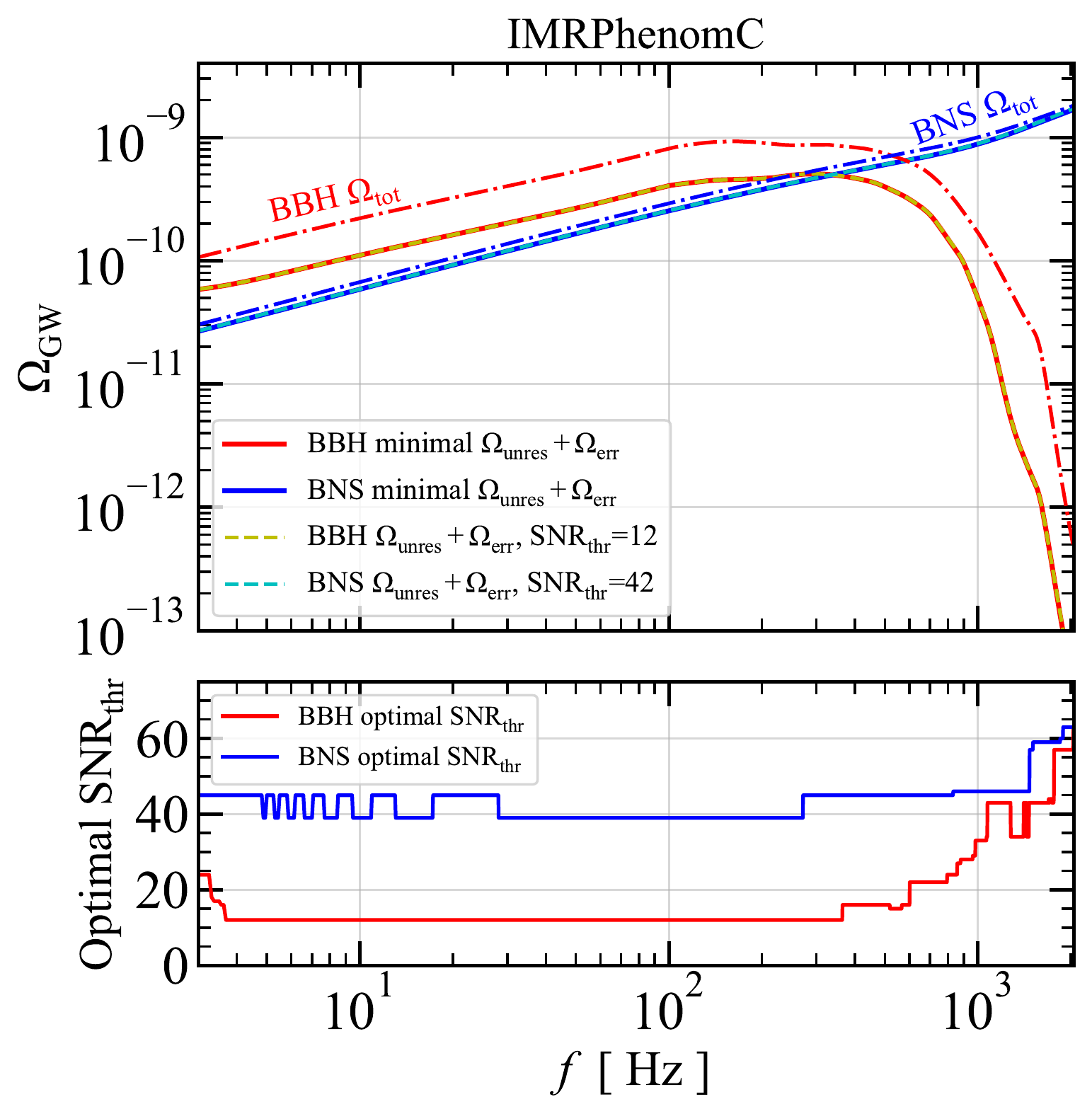}
\caption{Frequency-dependent minimized $\Omega_{\rm unres} + \Omega_{\rm err}$ for BBHs and BNSs (solid lines in the upper panels), and their corresponding optimal $\rm SNR_{thr}$ (lower panels), for two different waveform models: {\tt IMRPhenomD} (left) and {\tt IMRPhenomC} (right).  We also plot $\Omega_{\rm unres} + \Omega_{\rm err}$ for BBHs and BNSs computed by choosing the frequency-independent optimal $\rm SNR_{thr}$ that matches the results in the lower panel (dashed lines in the upper panels): the results are almost indistinguishable from the solid lines. For comparison, in the upper panels we also plot the total energy density $\Omega_{\rm tot}$ for BBHs and BNSs (dash-dotted lines). All results refer to our fiducial 3-detector network.}
\label{fig_freq_wise}
\end{figure*}

From a knowledge of the information matrix we can then draw the recovered parameters $\vec{\theta}^i_{\rm rec}$.
We consider a set of $9$ parameters for the calculation of the information matrix for each CBC event:
\begin{equation}
\vec{\theta} = 
\left\{ 
\ln(\frac{\mathcal{M}_z}{M_\odot}), 
\eta, 
\ln(\frac{D_L}{\text{Mpc}}), 
\cos\iota, 
\cos\delta,
\alpha, 
\psi,
\phi_{c}, 
t_{c} 
\right\} \,,
\label{eq_info_paras}
\end{equation}
where $\mathcal{M}_z = \mathcal{M} (1+z)$ is the detector-frame chirp mass.

\begin{figure}[h!]
\includegraphics[width=\columnwidth]{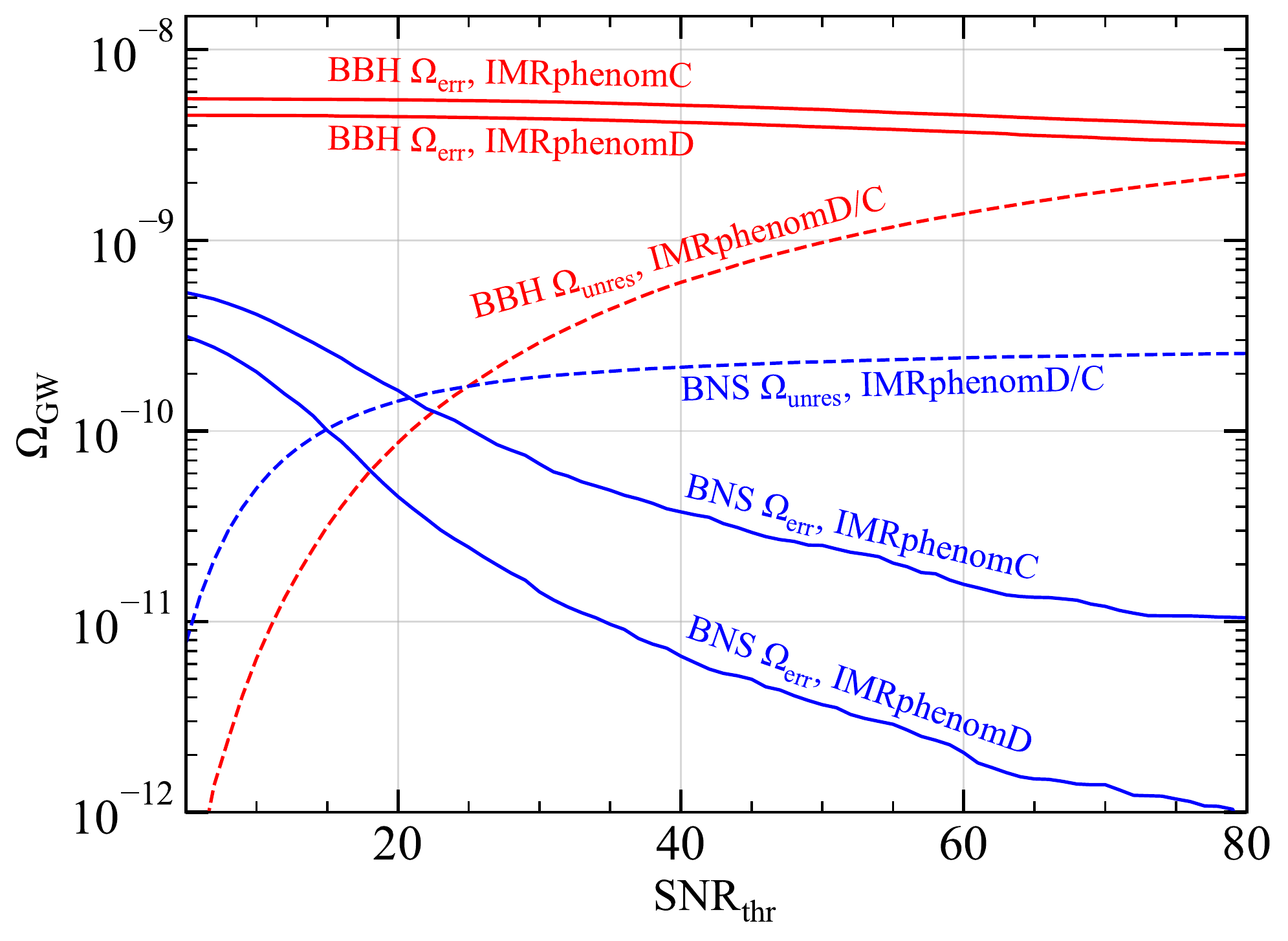}
\caption{Backgrounds $\Omega_{\rm err}$ (solid lines) and $\Omega_{\rm unres}$ (dashed lines) evaluated at $f=100$~Hz for our fiducial 3-detector network and for the two different waveform models, {\tt IMRPhenomD and }{\tt IMRPhenomC}. The choice of waveform model is irrelevant to determine the unresolved population $\Omega_{\rm unres}$ (compare Fig.~\ref{fig_fraction_resolved}), but it does affect $\Omega_{\rm err}$, in particular for BNSs. The difference in the BNS $\Omega_{\rm err}$ evaluated for different waveform models increases at large values of $\rm SNR_{thr}$, as expected.}
\label{fig_WFmodel}
\end{figure}

As mentioned in Sec.~\ref{sec_pop}, we ignore the 6 spin parameters because their effects are expected to be subdominant. 
We have reproduced all of the results in the previous study~\cite{Sachdev:2020bkk}, which considered only three parameters ($\mathcal{M}_z$, $\phi_c$, and $t_c$) when implementing the information-matrix formalism.
Compared to their work, our choice results in more realistic estimates of the statistical errors. 

In XG detectors, longer signals (e.g., close-by BNSs) are expected to overlap. The overlap has two main effects. In searches for GW signals, the presence of overlapping signals might cause some of them to be missed. Reference~\cite{Wu:2022pyg} estimates that this will reduce the redshift reach by 8\% for ET, and by 15\% for CE. In parameter estimation, it may be challenging to infer the parameters of multiple coincident signals simultaneously~\cite{Himemoto:2021ukb, Pizzati:2021apa, Samajdar:2021egv, Janquart:2022nyz}. Furthermore, multiple weak overlapping signals can produce a ``confusion noise'' contribution that must be added to instrumental noise, increasing the uncertainties on the recovered parameters of loud signals~\cite{Antonelli:2021vwg, Reali:2022aps}. Both effects (reduced redshift reach and larger parameter estimation errors) increase $\Omega_{\rm unres} + \Omega_{\rm err}$ compared to our estimates, which therefore can be considered conservative.

\subsection{Effect of waveform modeling systematics and choice of detector networks}
\label{sec_GWcalc_others}

The waveform polarizations $\tilde{h}_{+,\times}$ must be computed by assuming a specific model, which may lead to systematic biases that add further deviations in the recovered GW signals~\cite{Flanagan:1997kp,Miller:2005qu,Cutler:2007mi}. A detailed study of waveform systematics is beyond the scope of this work, but to roughly estimate their effect we show results for two different waveforms: we adopt {\tt IMRPhenomD}~\cite{Husa:2015iqa,Khan:2015jqa} as our ``fiducial'' reference model, and we compare it against {\tt IMRPhenomC}~\cite{Santamaria:2010yb} to investigate possible effects from modeling systematics.
We neglect tidal effects in BNSs, and we adopt the same waveform models for both our BNS and BBH populations. 

We study two possible networks of XG ground-based detectors. A multiple-detector network is needed because SGWBs are detected by cross-correlating the data of multiple detectors, with the background in one detector used as a matched filter for the data in the other detectors~\cite{Allen:1997ad, Schutz:1989cu}.
If the detectors are geographically well separated, which is our case, the risk of common noise of terrestrial origin is greatly reduced. The two network choices are:

\begin{itemize}
    \item A \emph{fiducial} scenario consisting of three XG observatories. We choose one CE detector with $40$-km arm length in the U.S., one CE with $20$-km arm length in Australia, and one ET in Italy. The locations of these detectors can be found in Table~III of Ref.~\cite{Borhanian:2020ypi} under the labels \texttt{C}, \texttt{S}, and \texttt{E}, respectively.    

    \item An \emph{optimistic} scenario with five XG observatories: four CE detectors with $40$-km arm length at the current locations of LIGO Hanford, Livingston, India and KAGRA, plus one ET at the location of Virgo. 
    The location choice is the same as the 5-detector case in previous studies~\cite{Regimbau:2016ike, Sachdev:2020bkk}, to facilitate a direct comparison of the results.
    The coordinates for the detectors in this network can be found under the labels \texttt{H}, \texttt{L}, \texttt{I}, \texttt{K}, and \texttt{V} in Table~III of Ref.~\cite{Borhanian:2020ypi}.     
\end{itemize}

We chose networks of three and five XG detectors to facilitate comparison with recent studies~\cite{Regimbau:2016ike, Sachdev:2020bkk, Borhanian:2022czq, Iacovelli:2022bbs, Reali:2022aps}.
The detection SNR and the statistical errors on the recovered parameters depend on the detector network considered: see Eqs.~\eqref{eq_snrnet} and \eqref{eq_infonet}. 

Figure~\ref{fig_psds} shows the power spectral density, or more precisely $\sqrt{S_n(f)}$, of these detectors. 
We assume all the CE detectors to be in the latest phase of their development and optimized at low frequencies, i.e., we select the \texttt{CE2} and the \texttt{CBO} option in {\tt GWBENCH}~\cite{Borhanian:2020ypi}. Here CE2 stands for the final stage of CE detectors (in contrast to CE1, which means the initial stage) and CBO stands for ``compact-binary optimizations'' (in contrast to PMO, which means ``post-merger optimizations''). For ET, we consider its most up-to-date triangular configuration with ET-D sensitivity~\cite{Punturo:2010zz}. 

We set the minimum frequency in our calculations to be $3~\rm{Hz}$, which is consistent with the PSDs of the detectors in our networks (Fig.~\ref{fig_psds}).
Regarding the final frequency, we use $2048~\rm{Hz}$. This is larger than the frequency at the innermost stable circular orbit~\cite{Misner:1973prb} for most of the sources in our catalog, and the detector noise makes contributions above this maximum frequency negligible~\cite{Borhanian:2022czq}.

We include Earth-rotation effects for BNSs and neglect them for BBHs. 
This is because GW signals from BBHs last only a few minutes in the detector frame~\cite{Pizzati:2021apa}, during which the change in the detectors' response~\cite{Chan:2018csa} is small, so the effect on their SNR and parameter estimation is negligible~\cite{Borhanian:2022czq}. For BNSs, on the contrary, the signals last up to several hours~\cite{Pizzati:2021apa}, and hence the change of the detectors' response must be taken into account~\cite{Borhanian:2022czq}.

\begin{figure*}
\includegraphics[width=\columnwidth]{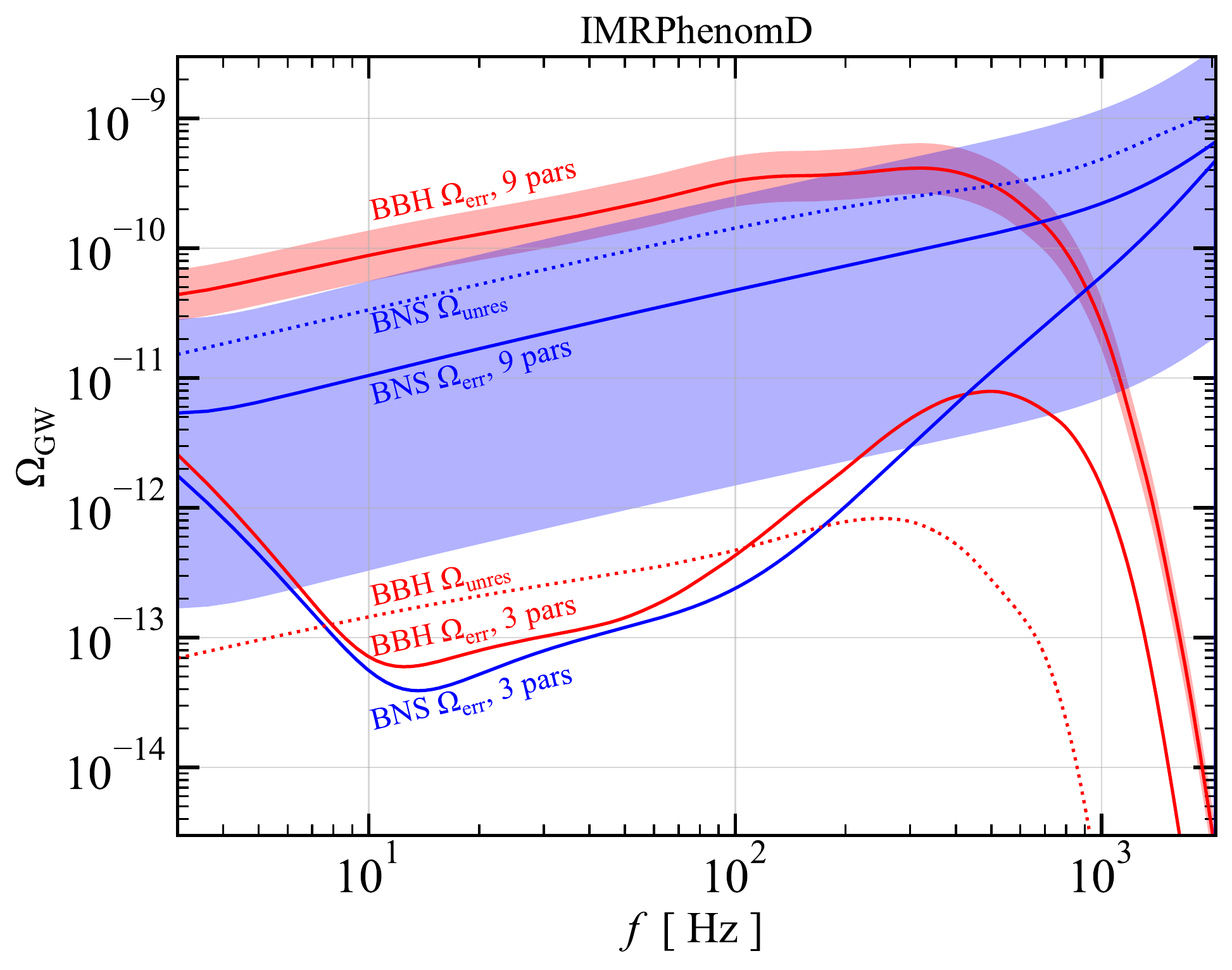}
\includegraphics[width=\columnwidth]{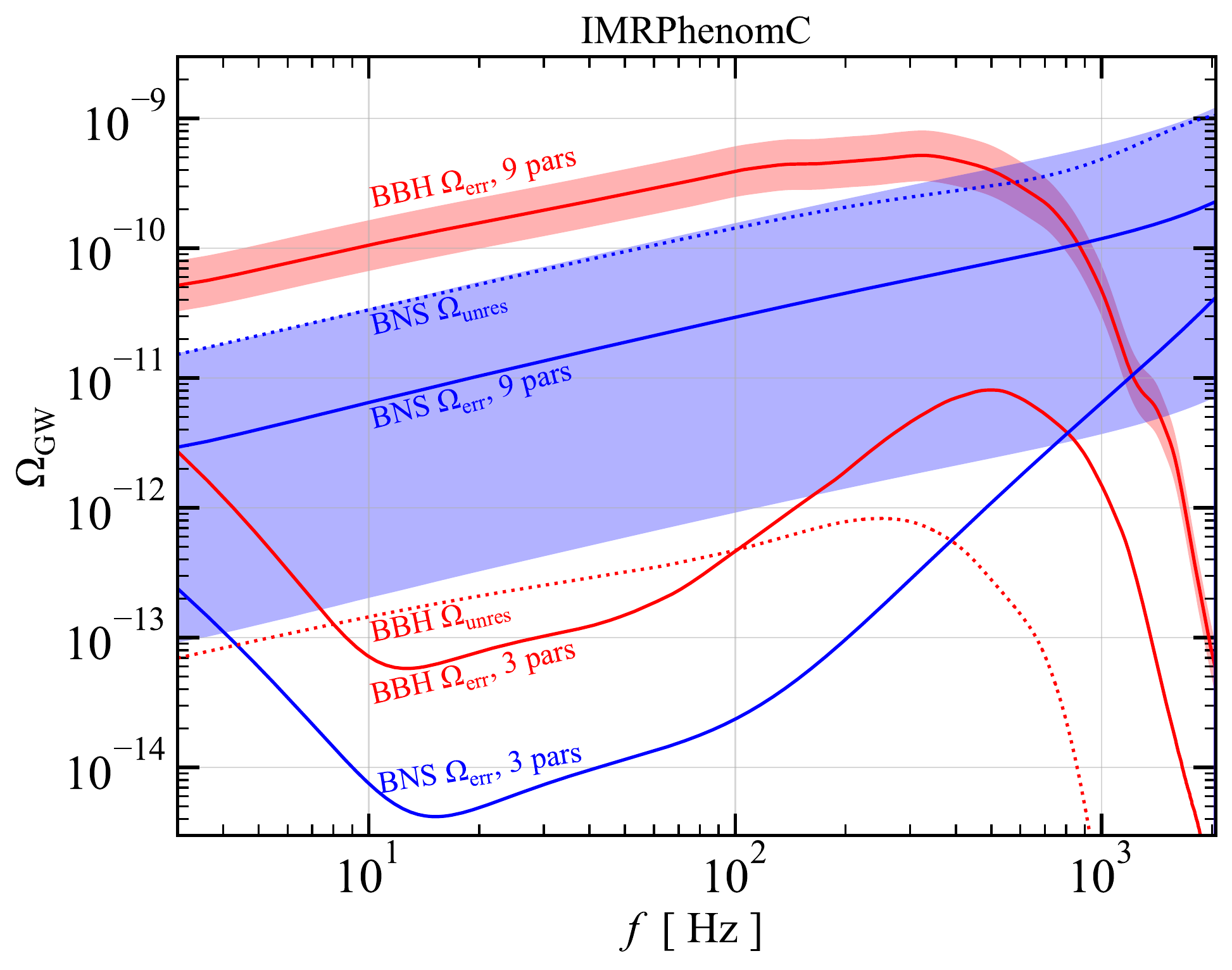}
\caption{
Unresolved background $\Omega_{\rm unres}$ (dotted lines) and error contributions to the background $\Omega_{\rm err}$ (solid lines) computed using both our 9-parameter recovery (``9 pars'') and the 3-parameter recovery (including only $\mathcal{M}_z$, $t_c$, and $\phi_c$) considered in previous work~\cite{Sachdev:2020bkk} (``3 pars'').  
All backgrounds are computed at the frequency-independent optimal $\rm SNR_{thr}$ for the 9-parameter case: 10 (for BBHs) and 20 (for BNSs) in the left panel; 12 (for BBHs) and 42 (for BNSs) in the right panel.
The shaded band around the 9-parameter $\Omega_{\rm err}$  shows astrophysical uncertainties on the rates. The worsening in $\Omega_{\rm err}$ due to including 9 parameters instead of 3 is quite dramatic and it is larger than astrophysical uncertainties, especially for BBHs.
} 
\label{fig_OmegaErr_3vs9}
\end{figure*}

\begin{figure*}[t!]
\includegraphics[width=\columnwidth]{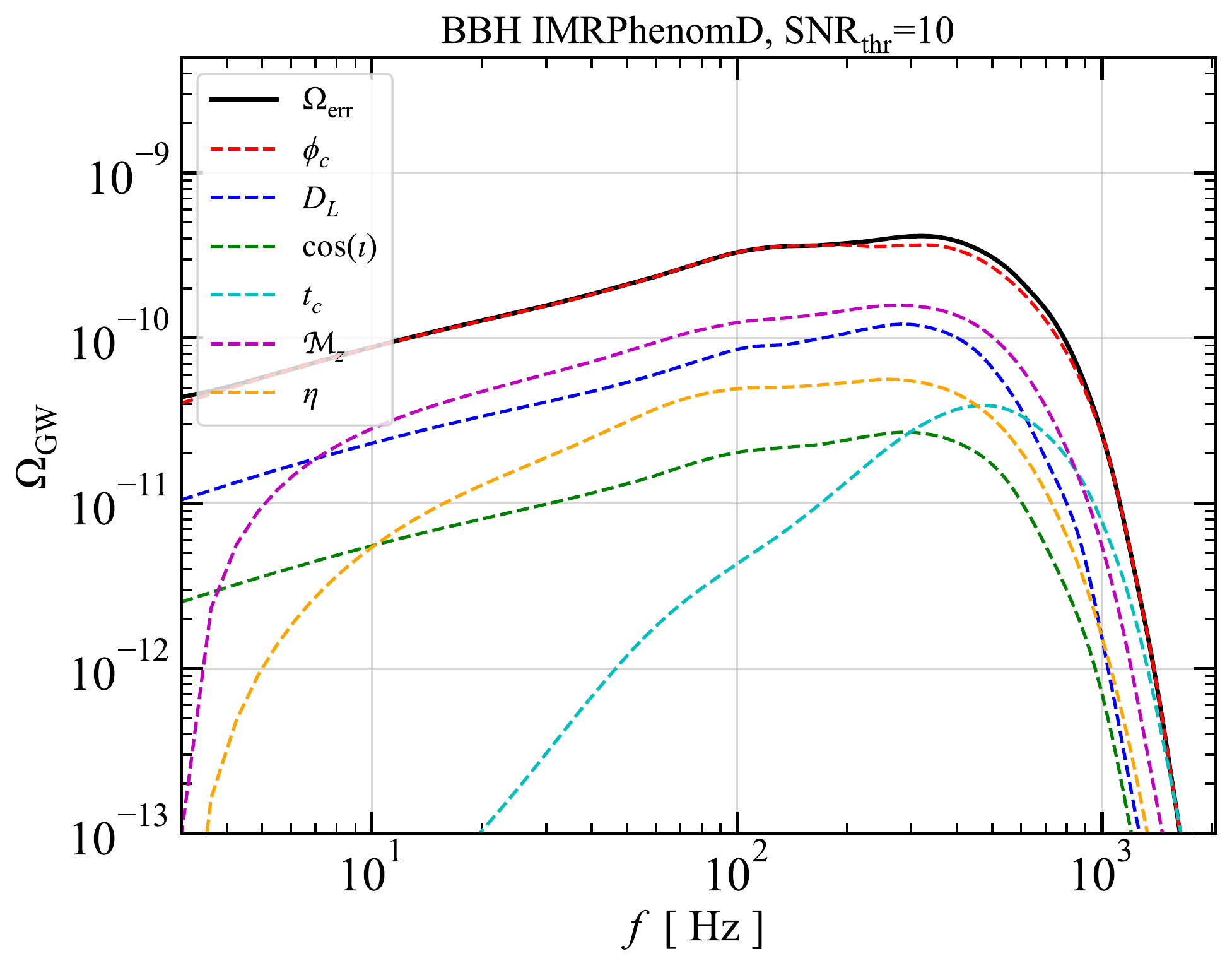}
\includegraphics[width=\columnwidth]{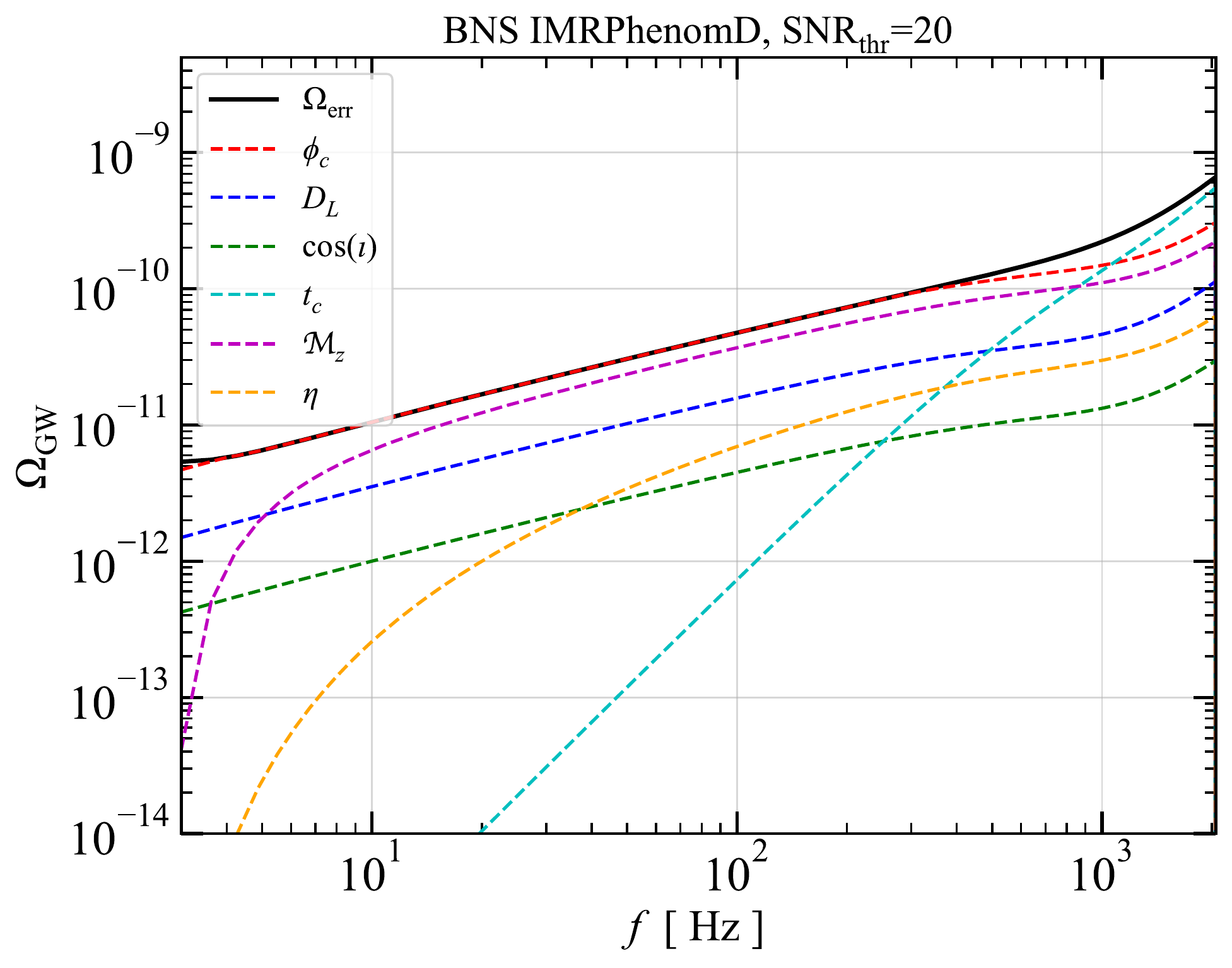}
\includegraphics[width=\columnwidth]{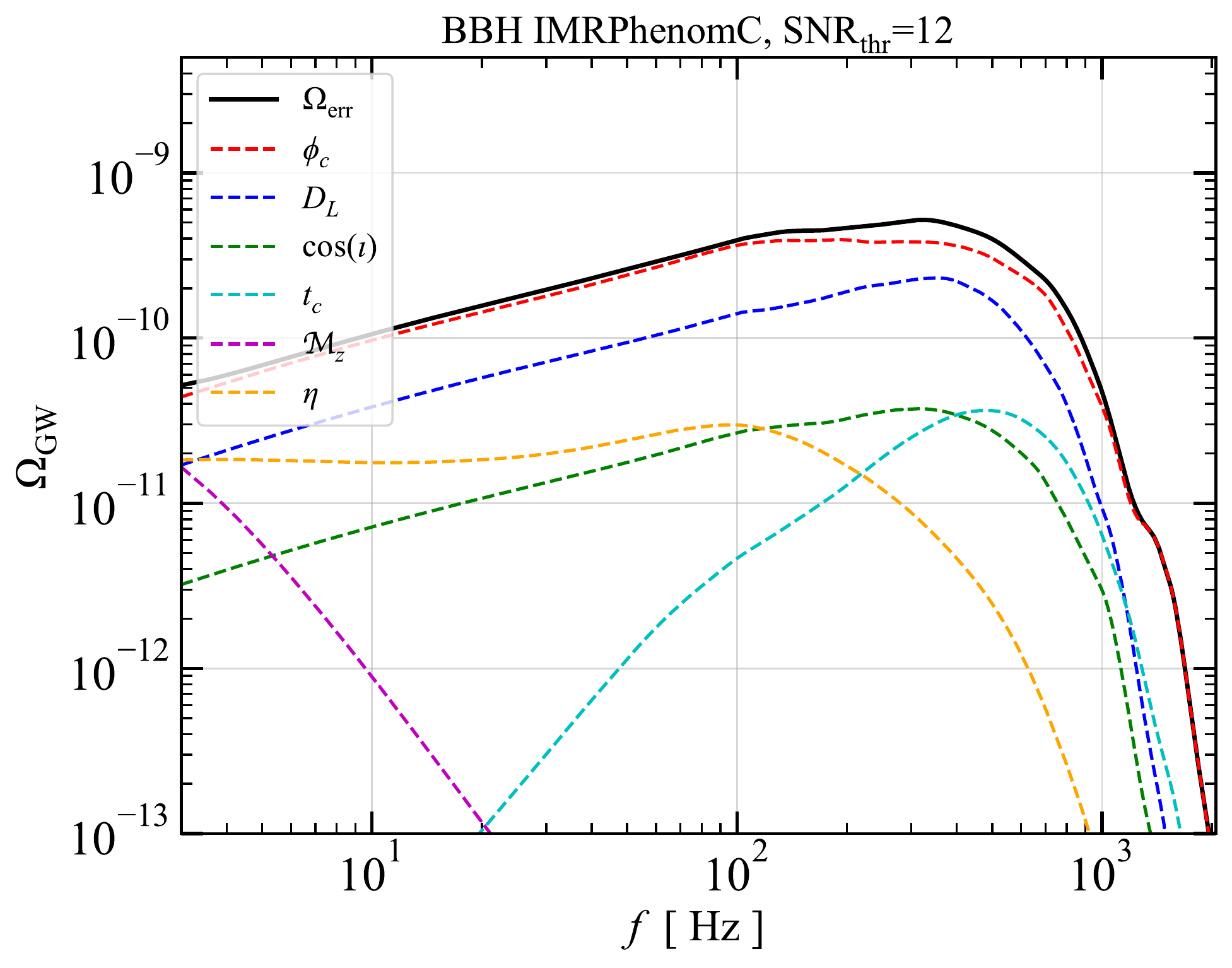}
\includegraphics[width=\columnwidth]{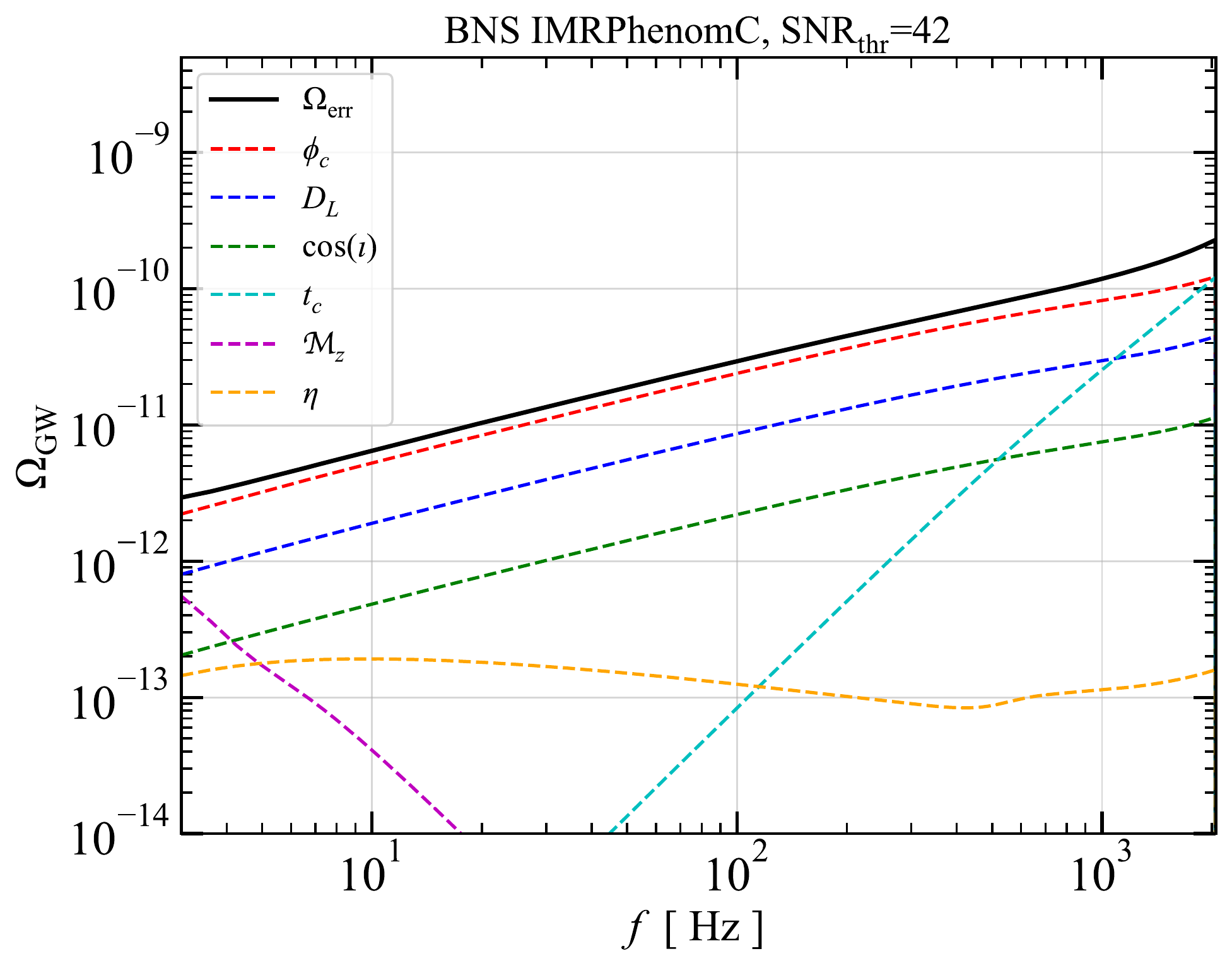}
\caption{Breakdown of $\Omega_{\rm err}$ (at the optimal SNR thresholds) showing the contribution from the errors on each of the parameters.
All results are for our fiducial 3-detector network.
}
\label{fig_Omegas_err}
\end{figure*}

\section{Results and discussion}
\label{sec_rslt}

In this section, we present and discuss our results. We mainly focus on our fiducial 3-detector network.
We usually compare our fiducial waveform model {\tt IMRPhenomD} to the older model {\tt IMRPhenomC} as a simple way to quantify waveform modeling uncertainties.

Figure~\ref{fig_Omegas} shows $\Omega_{\rm tot}$, $\Omega_{\rm unres}$, and $\Omega_{\rm err}$ for BBHs and BNSs, computed using different values of $\rm SNR_{thr}$ (as indicated in the legend) and the two waveform models. 
For BBHs, the whole signal (inspiral, merger, and ringdown) can be observed in the frequency range of interest. For BNSs, by contrast, the inspiral phase dominates. In both cases, we recover the expected inspiral behavior at low frequencies, where $\Omega_{\rm tot}\propto \Omega_{\rm unres} \propto f^{2/3}$~\cite{Maggiore:1999vm}.

For all the panels in Fig.~\ref{fig_Omegas}, as we increase $\rm SNR_{thr}$ the number of unresolved CBC events increases, and so does $\Omega_{\rm unres}$. On the contrary, $\Omega_{\rm err}$ decreases with growing $\rm SNR_{thr}$, because the number of resolved CBC events decreases and the events with larger SNRs have smaller parameter estimation errors.

By comparing the left and right panels we see that the difference between {\tt IMRPhenomD} and {\tt IMRPhenomC} is visible but small, especially for BBHs.
The difference between $\Omega_{\rm err}$ for PhenomD and PhenomC becomes more noticeable as $\rm SNR_{thr}$ grows, because differences between waveforms dominate over statistical errors for large-SNR signals.

Figure~\ref{fig_fraction_resolved} shows the fraction of resolved events as a function of $\rm SNR_{thr}$.
For BBHs, 99\%, 53\%, and 20\% of the total events are resolved above $\rm SNR_{thr} =$ 8, 40, and 80, respectively.
Resolving BNSs is harder: only 29\%, 8\%, and 1\% of the total events are resolved above $\rm SNR_{thr} =$ 12, 20, and 40, respectively.
These numbers do not depend on the waveform model.

For small values of $\rm SNR_{thr}$ (8 or 12) most events are detected, so the unresolved contribution to the background is small, in the sense that $\Omega_{\rm unres} \ll \Omega_{\rm err}$ and $\Omega_{\rm unres}\ll \Omega_{\rm tot}$. On the contrary, for large $\rm SNR_{thr}$ (say, $\rm SNR_{thr}=80$ for BBHs), $\Omega_{\rm err}$ and $\Omega_{\rm unres}$ become comparable and overwhelm the detectors' sensitivity. As we increase $\rm SNR_{thr}$, $\Omega_{\rm unres}+\Omega_{\rm err}$ first decreases and then increases.
If we want to search for SGWBs of cosmological origin or for subdominant astrophysical SGWBs, our goal is to minimize $\Omega_{\rm unres}+\Omega_{\rm err}$. This happens at some \textit{optimal value of $\rm SNR_{thr}$} that we wish to determine.  

Figure~\ref{fig_freq_wise} shows our frequency-dependent, minimized $\Omega_{\rm unres} + \Omega_{\rm err}$ for the two subpopulations of BBHs and BNSs (upper panel), as well as the corresponding \textit{optimal $\rm SNR_{thr}$}, which in general is also a function of frequency (lower panel). 
It is clear from this plot that the optimal $\rm SNR_{thr}$ is nearly constant below hundreds of Hz, with only small fluctuations. This is because $\Omega_{\rm unres} + \Omega_{\rm err}$ varies very slowly with $\rm SNR_{thr}$, especially near the optimal $\rm SNR_{thr}$. Above hundreds of Hz, where merger and ringdown dominate, the optimal $\rm SNR_{thr}$ increases. This effect is more prominent for BBHs, whose energy density drops rapidly in this frequency range.

The mild variability of $\rm SNR_{thr}$ with frequency means that we can compute $\Omega_{\rm unres} + \Omega_{\rm err}$ using a frequency-independent optimal $\rm SNR_{thr}$. We choose this optimal $\rm SNR_{thr}$ to be 10 (20) for BBHs (BNSs) when we consider the {\tt IMRPhenomD} model, and 12 (42) for BBHs (BNSs) when we work with the {\tt IMRPhenomC} model. These values are chosen to match the frequency-dependent optimal $\rm SNR_{thr}$ computed in the lower panel.
The dashed lines in the upper panel of Fig.~\ref{fig_freq_wise} show that the minimal $\Omega_{\rm unres} + \Omega_{\rm err}$ computed using a frequency-independent optimal $\rm SNR_{thr}$ is nearly identical to the value computed using a frequency-dependent detection threshold. This is very convenient from an experimental point of view.
In the rest of this section we will compute $\Omega_{\rm unres} + \Omega_{\rm err}$ using a frequency-independent optimal $\rm SNR_{thr}$.

Figure~\ref{fig_WFmodel} clarifies why the optimal $\rm SNR_{thr}$ is so different for the two waveform models in the BNS case, while it's not for BBHs.
The main effect of a changing $\rm SNR_{thr}$ is to uniformly shift $\Omega_{\rm unres}$ and $\Omega_{\rm err}$ up or down at any given frequency, so (for concreteness) in Fig.~\ref{fig_WFmodel} we evaluate all backgrounds at a single frequency ($f=100$~Hz), and we plot them as functions of $\rm SNR_{thr}$ for different waveform models.  The choice of waveform model does not affect $\Omega_{\rm unres}$ (compare Fig.~\ref{fig_fraction_resolved}), but it does affect $\Omega_{\rm err}$. In particular, the BNS estimates for $\Omega_{\rm err}$ evaluated using {\tt IMRPhenomD} and {\tt IMRPhenomC} are remarkably different. This difference increases at large values of $\rm SNR_{thr}$, as expected, because systematic errors dominate over statistical uncertainties in this regime.  The large variation in $\Omega_{\rm err}$ for BNSs explains why the optimal $\rm SNR_{thr}$ that minimizes the sum $\Omega_{\rm unres} + \Omega_{\rm err}$ can vary by as much as a factor of order 2, from $20$ (for BNS backgrounds computed using {\tt IMRPhenomD}) to $42$ (for BNS backgrounds computed using {\tt IMRPhenomC}). The larger difference in $\Omega_{\rm err}$ for BNSs must lie in the differences in how {\tt IMRPhenomC} and {\tt IMRPhenomD} treat the inspiral part of the signal, because BNS signals are inspiral-dominated. In both models, the inspiral phase in the frequency domain is modeled using a post-Newtonian (PN) expansion, i.e. {\tt Taylorf2}~\cite{Sathyaprakash:1991mt,Buonanno:2009zt}, plus higher-order corrections tuned to inspiral-merger-ringdown (IMR) hybrids~\cite{Santamaria:2010yb,Khan:2015jqa}. However, in {\tt IMRPhenomC} the {\tt Taylorf2} model is also used for the inspiral phase of the IMR hybrids, while in {\tt IMRPhenomD} the inspiral phase of the IMR hybrids is constructed with an effective-one-body model, {\tt SEOBNRv2}~\cite{Taracchini:2012ig}. Further uncertainties related to waveform systematics may be caused by the inclusion of additional PN terms, such as tidal deformations. Note that uncertainties due to waveform systematics are subdominant compared to other uncertainties (e.g., on the local merger rates) considered here. A thorough analysis of waveform systematics is beyond the scope of this paper. 

Figure~\ref{fig_OmegaErr_3vs9} shows $\Omega_{\rm unres}$ and $\Omega_{\rm err}$ for BBHs and BNSs computed at the optimal $\rm SNR_{thr}$ with the two waveform models. For $\Omega_{\rm err}$, we show the comparison between our calculation, in which we include 9 parameters, and the 3-parameter estimate of $\Omega_{\rm err}$ (which included only $\mathcal{M}_z$, $t_c$, $\phi_c$) considered in previous work~\cite{Sachdev:2020bkk}. 
Our 3-parameter estimate of $\Omega_{\rm err}$ is similar in amplitude and functional form to the results of Ref.~\cite{Sachdev:2020bkk}: in fact, we checked that we can recover their results if we use the same value of $\rm SNR_{thr}$ and the same population model. 
The difference between the residual backgrounds $\Omega_{\rm err}$ computed using our full 9-parameter recovery and those computed using the 3-parameter recovery of Ref.~\cite{Sachdev:2020bkk} is quite striking. The 9-parameter estimate of $\Omega_{\rm err}$ is 2--3 orders of magnitude larger than the 3-parameter estimate of $\Omega_{\rm err}$ at the same $\rm SNR_{thr}$. As we elaborate below, there are two reasons for this difference. One is quite obvious: as we increase the number of parameters, correlations and degeneracies between different parameters significantly increase the spread of the posteriors, and therefore the typical deviations of the recovered parameters from their true values become much larger. The second reason is not so obvious: some of the parameters that were not included in Ref.~\cite{Sachdev:2020bkk} affect directly the estimate of the signal’s amplitude, and signals with poorly known amplitude have a large impact on the removal of the foreground.

The shaded bands in Fig.~\ref{fig_OmegaErr_3vs9} show the impact of astrophysical uncertainties on the local merger rates (note that this is a lower bound on astrophysical uncertainties, because the redshift evolution of the rates is even more poorly constrained). It is difficult to formulate reliable predictions for the CBC SGWB, especially for BNSs, where the background can vary by about two orders of magnitude. However, the increase in $\Omega_{\rm err}$ due to the addition of the amplitude parameters is even larger than the variability of the background due to the uncertain merger rates. There is a clearly visible difference in the residual background $\Omega_{\rm err}$ predicted by the two waveform models, which is larger for BNSs than for BBHs. This difference, while small compared to current astrophysical uncertainties, highlights the importance of waveform systematics in data analysis and it will play a more prominent role in SGWB forecasts in the coming years, as new detections will steadily reduce the uncertainties in merger rates.

We can better understand why the 9-parameter estimate of $\Omega_{\rm err}$ is so much larger than the 3-parameter estimate by identifying which parameters have large uncertainties, and therefore can give a large contribution to $\Omega_{\rm err}$ even when $\rm SNR_{thr}$ is large.
To compute the contribution to $\Omega_{\rm err}$ due to each parameter, we evaluate Eq.~\eqref{eq_Omega_err} by using the recovered value {\em of that parameter only} in $\vec{\theta}^i_{\rm rec} $, while we use the ``true'' values for all other parameters.
We perform this calculation on each of the parameters in Eq.~\eqref{eq_info_paras}, with three exceptions: $\alpha$, $\delta$, and $\psi$. The reason is that these three parameters do not affect the waveform polarizations (hence the energy flux and $\Omega$), but only the detectors' response (hence the SNR and the information matrix). However, they have a very important indirect impact on $\Omega_{\rm err}$ through their correlations with the other six parameters.

Figure~\ref{fig_Omegas_err} shows the results for BBHs (left) and BNSs (right), using either {\tt IMRPhenomD} (top) or {\tt IMRPhenomC} (bottom). In each of the panels, we compute the errors at the optimal value of $\rm SNR_{thr}$. One common feature of all panels is that large contributions to the uncertainty at nearly all frequencies arise from $\phi_c$ and $D_L$, with $\phi_c$ generally yielding the dominant contribution. 
The dominance of the error on $\phi_c$ is mostly caused by its degeneracy with the polarization angle $\psi$, as both parameters contribute to an overall phase term in the detector response (see e.g.~\cite{Berti:2004bd,Buonanno:2007yg}). The degeneracy is more severe for nearly face-on/face-off binaries, because for these binaries a rotation about the line of sight is identical to a shift in the orbital phase~\cite{Buonanno:2007yg,Maggiore:2007ulw}. These systems also generate louder signals, meaning that detected sources tend to be preferentially close to face-on/face-off binaries~\cite{Nissanke:2009kt,Schutz:2011tw}. 
The luminosity distance is highly degenerate with the inclination $\iota$, since both parameters appear only in the amplitudes of the waveform polarizations at leading order (see, e.g., Refs.~\cite{Maggiore:2007ulw,Colpi:2016fup}). In order to break this degeneracy, one would need to determine the two amplitudes of the $+$ and $\times$ polarizations independently, but the difference between these two amplitudes is quite small for most binary systems~\cite{Usman:2018imj}. 
In fact, various studies in the literature have shown that the luminosity distance remains poorly constrained for a significant subset of high-redshift, low-SNR sources, even with a network of XG detectors~\cite{Borhanian:2022czq,Ronchini:2022gwk,Iacovelli:2022bbs}.

\begin{figure*}[t!]
\includegraphics[width=0.32\textwidth]{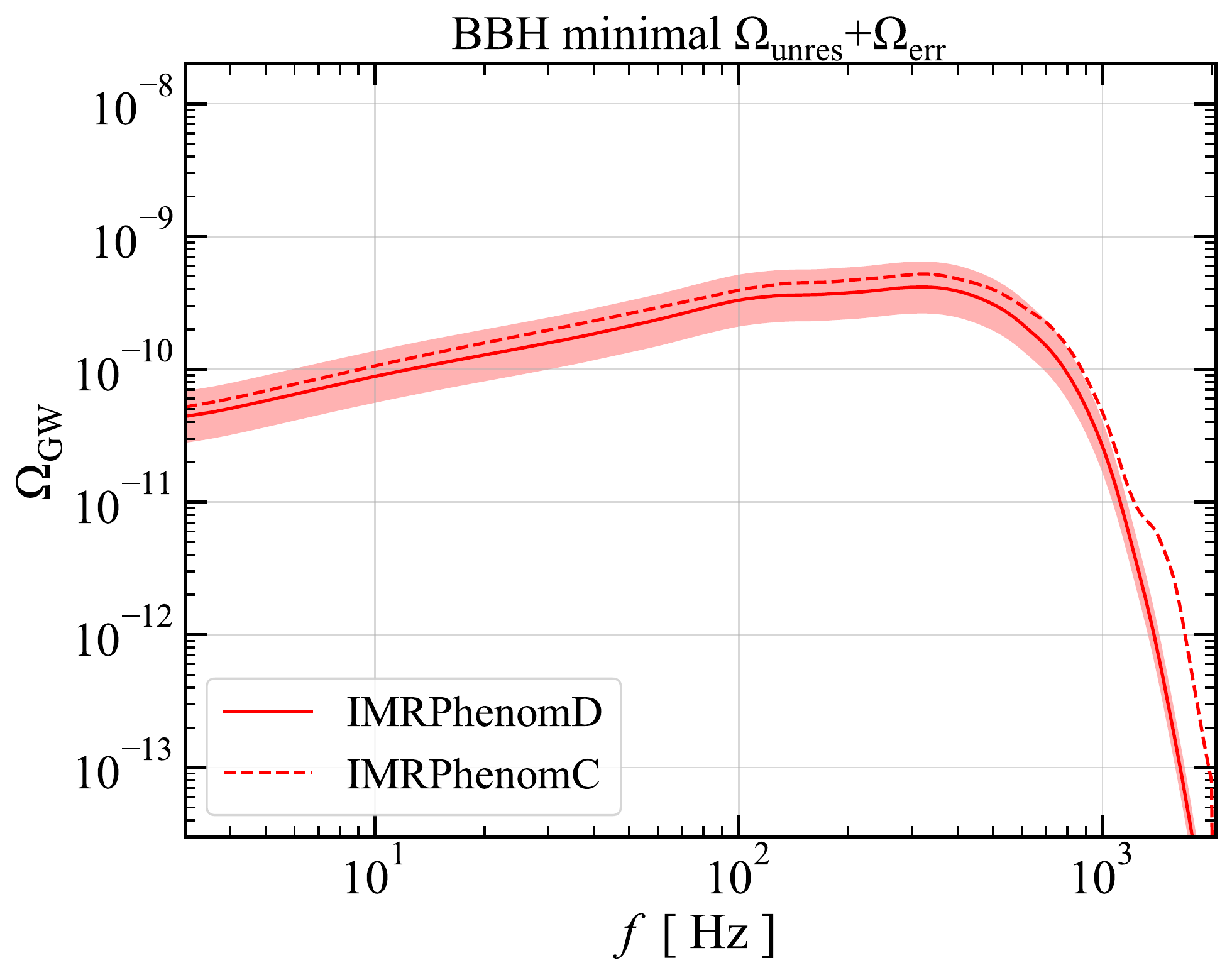}
\includegraphics[width=0.32\textwidth]{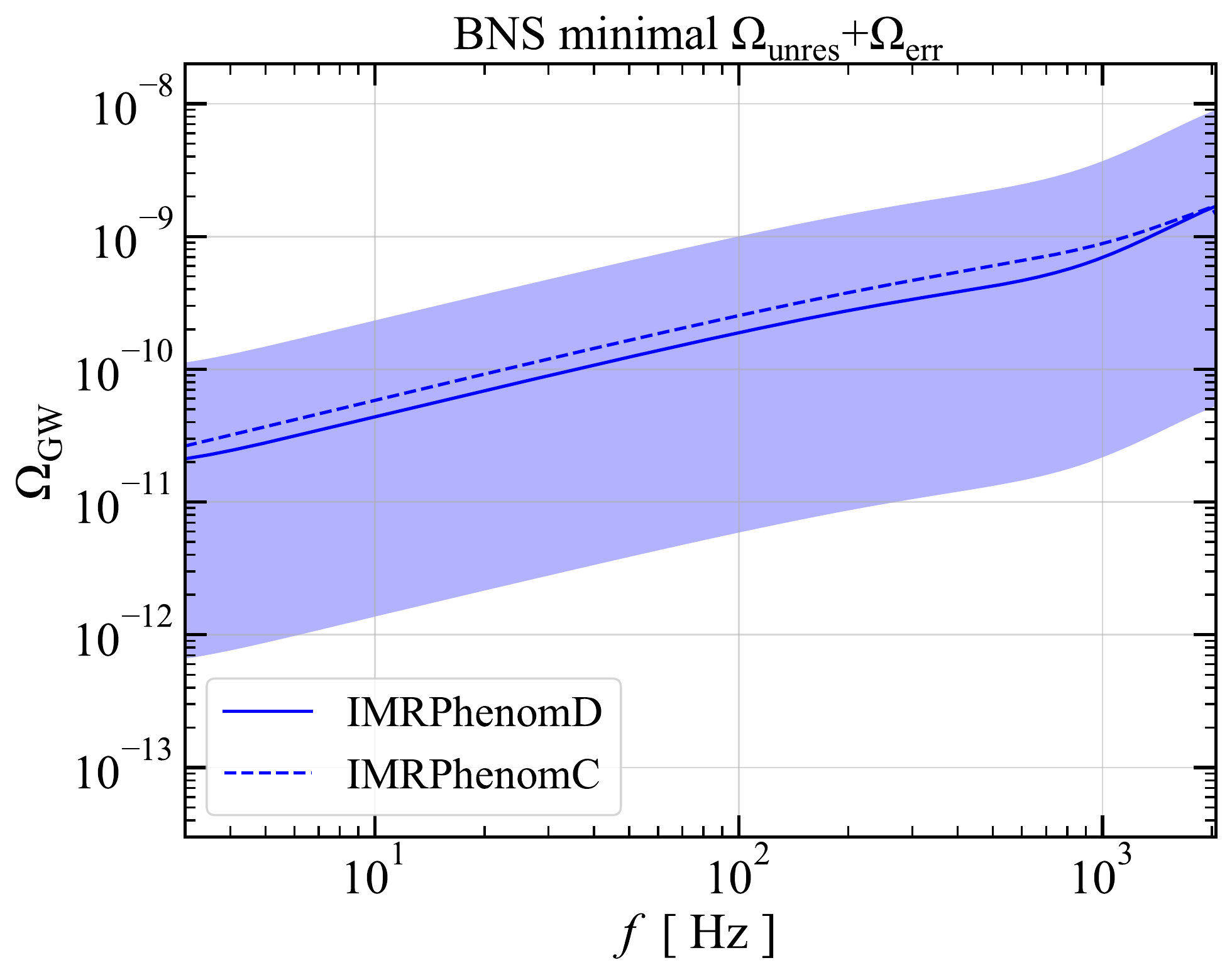}
\includegraphics[width=0.32\textwidth]{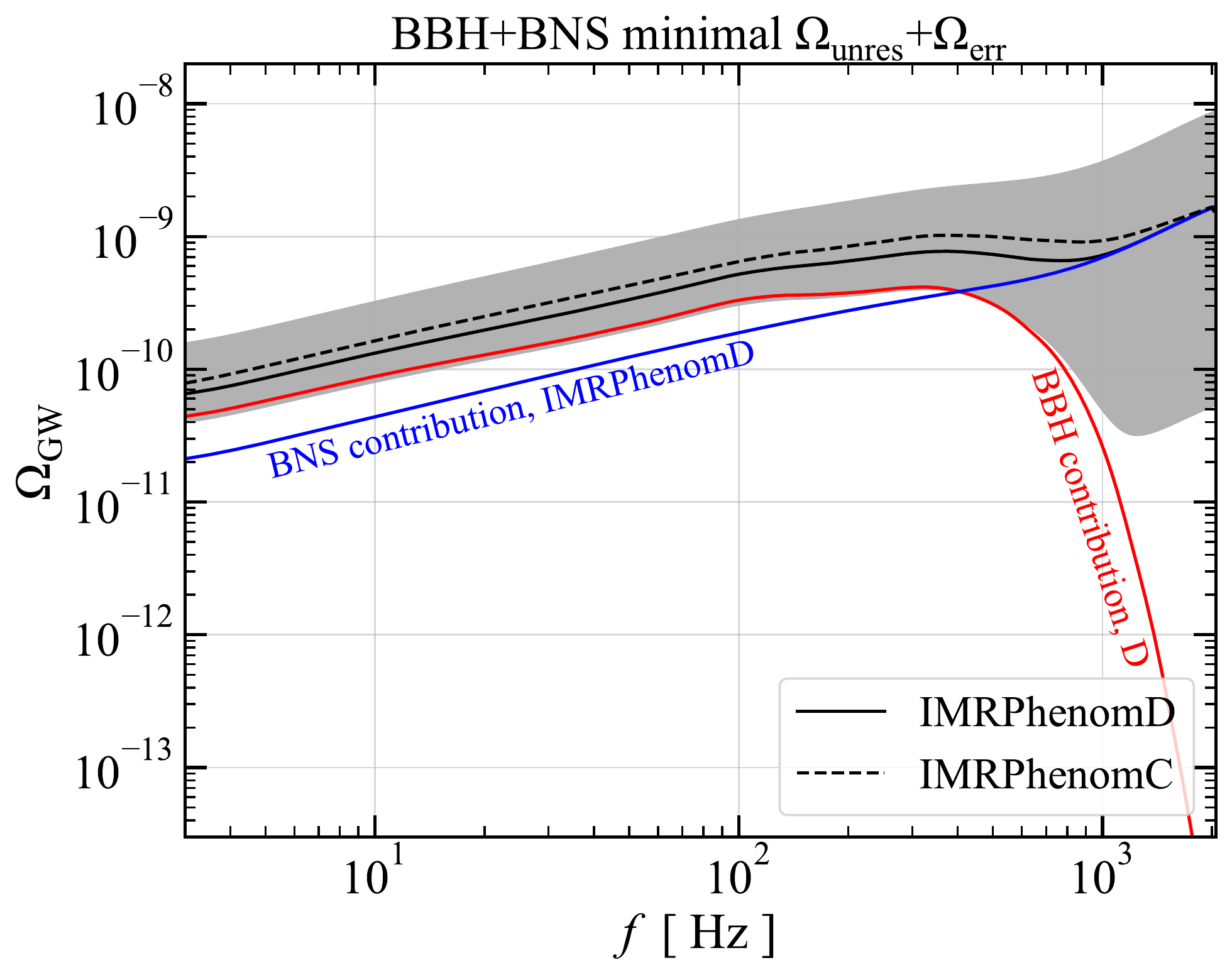}
\caption{Minimal value of $\Omega_{\rm unres} + \Omega_{\rm err}$ (i.e., ``total'' unresolved background corresponding to the optimal $\rm SNR_{thr}$) including astrophysical rate uncertainties for BBHs (left), BNSs (center) and for the combined populations (right). Solid lines refer to {\tt IMRPhenomD}, dashed lines to {\tt IMRPhenomC}. In each panel, for clarity, the uncertainty band at 90\% confidence level is shown only for {\tt IMRPhenomD}; the backgrounds computed for {\tt IMRPhenomC} have the same relative uncertainty.
The error band in the right panel is larger than that in the middle panel: it only looks narrower because we are using a log scale for the $y$ axis.
The red and blue solid curves in the right panel represent the contributions from BBHs and BNSs. They are the same as in the left and middle panels, but we show them together to facilitate comparisons.
All results refer to our fiducial 3-detector network.
}
\label{fig_Omega_plus_uncert}
\end{figure*}

In principle, these degeneracies can be broken or at least reduced by adopting waveform models that include higher harmonics~\cite{OShaughnessy:2014shr,Arun:2007hu,Arun:2007qv,Lang:2011je,Porter:2008kn}. However, contributions from higher harmonics are expected to be significant only for binaries whose components have fairly different masses~\cite{Cotesta:2018fcv}, while the majority of the systems in our catalogs are nearly equal mass. We do not expect the inclusion of higher modes to significantly reduce $\Omega_{\rm err}$ (especially in the BNS case), and we leave a more detailed exploration of higher harmonics to future work.

The main take-home message of this discussion is that including the contribution of parameters that affect the waveform amplitudes (such as $\iota$ and $D_L$) and the antenna patterns (such as $\alpha$, $\delta$ and $\psi$) is crucial to understand whether we can remove compact binary foregrounds.

Interestingly, the two waveform models {\tt IMRPhenomD} and {\tt IMRPhenomC} show large differences in the contribution from the two mass parameters ($\mathcal{M}_z$ and $\eta$) to $\Omega_{\rm err}$. In particular, the $\mathcal{M}_z$ contribution to $\Omega_{\rm err}$ is mostly negligible at all frequencies for {\tt IMRPhenomC}, while it becomes even larger than the contribution from the luminosity distance for {\tt IMRPhenomD}. Careful scrutiny shows that this is not due to differences in the information matrices, which are similar for the two waveform models. The effect is more subtle, and related to waveform systematics: small perturbations in $\mathcal{M}_z$ and $\eta$ in the two models are different, and their effect piles up when we consider the whole population. We have tested this observation by using exactly the same information matrices to sample $\vec{\theta}^i_{\rm rec}$ in Eq.~\eqref{eq_Omega_err}: even in this case we get different results for {\tt IMRPhenomD} and {\tt IMRPhenomC}, which are qualitatively similar to the results shown in Fig.~\ref{fig_Omegas_err}. The contributions due to the $\mathcal{M}_z$ and $\eta$ parameters are always subdominant compared to other parameters (in particular the coalescence phase), and therefore these differences have a small impact on the overall estimate of $\Omega_{\rm err}$, but they do highlight the fact that waveform systematics will play an important role in the data analysis of large populations in the XG detector era.

While the effect of waveform systematics is important, its effect on the estimate of $\Omega_{\rm unres}+\Omega_{\rm err}$ is currently dwarfed by astrophysical rate uncertainties (Sec.~\ref{sec_pop_rate}). Figure~\ref{fig_Omega_plus_uncert} compares the uncertainties of our calculated minimal $\Omega_{\rm unres} + \Omega_{\rm err}$ due to the two different waveform models with astrophysical rate uncertainties for BBHs (left panel), BNSs (central panel), and their sum (right panel). The uncertainty band in each panel is shown for {\tt IMRPhenomD} only ({\tt IMRPhenomC} has the same relative uncertainty). The difference between the two waveform models is negligible compared to astrophysical rate uncertainties, the only exception being BBHs at the highest frequencies, where the energy density is low and the merger-ringdown part of the waveform dominates the signal. By looking at the total $\Omega_{\rm unres}+\Omega_{\rm err}$ found by adding BBHs and BNSs (right panel) we see that BBHs dominate below $\simeq 400$~Hz, while BNSs dominate at higher frequencies.

\begin{figure}[t!]
\includegraphics[width=\columnwidth]{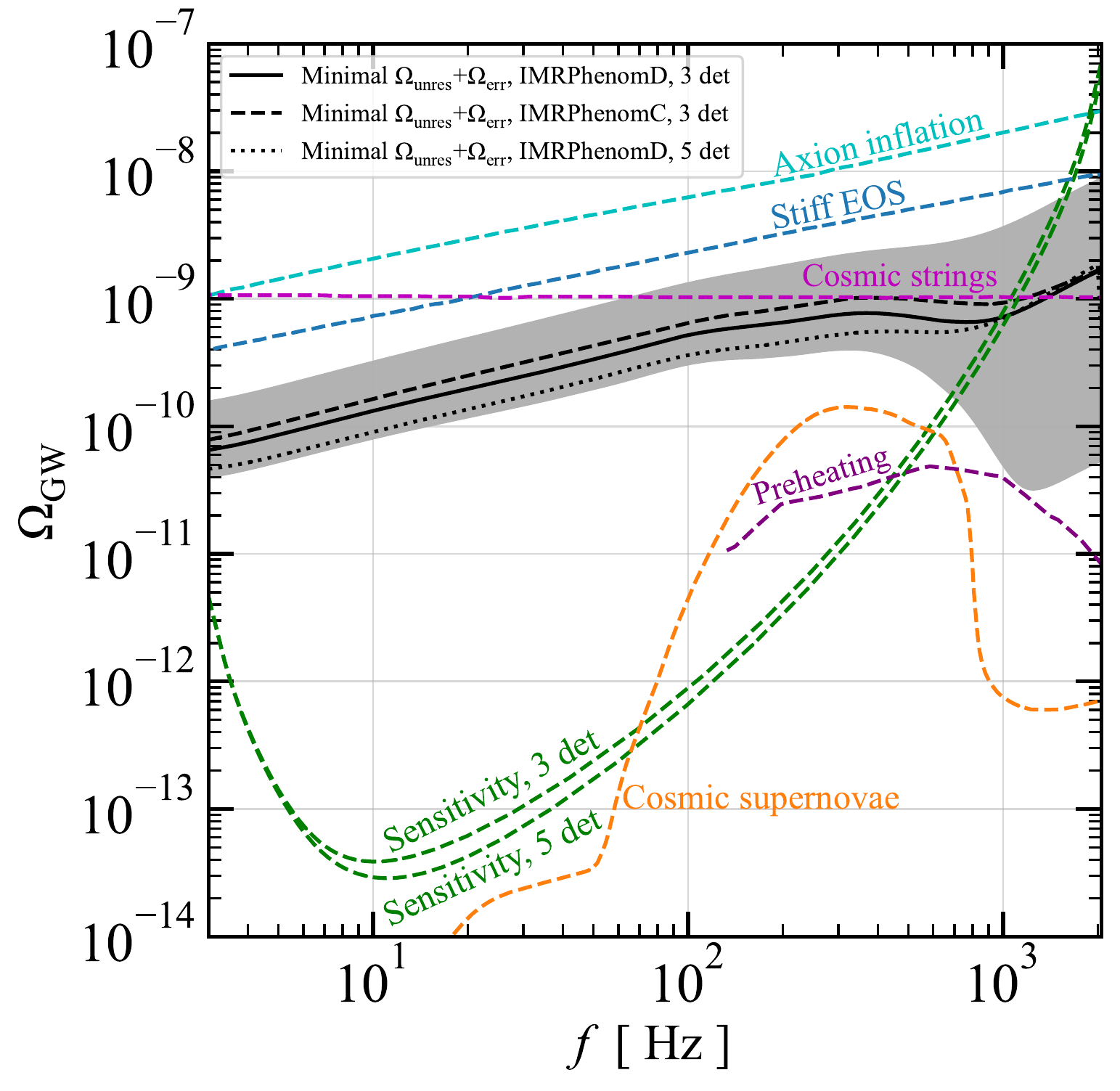}
\caption{
Minimal value of $ \Omega_{\rm unres} + \Omega_{\rm err} $ for BBHs and BNSs combined (black lines+grey band). This quantity is a good estimate of the total CBC foreground when observing the SGWB produced by other possible sources, as labeled. For our fiducial 3-detector network, we show results for both {\tt IMRPhenomD} (solid) and {\tt IMRPhenomC} (dashed). We also show {\tt IMRPhenomD} results for the optimistic 5-detector network (dotted).
The astrophysical uncertainty band at 90\% confidence level is shown only in our fiducial case of {\tt IMRPhenomD} and 3 detectors, for clarity.
Green dashed lines are the sensitivities of XG detectors to SGWBs, assuming 1-year integration, in the absence of foregrounds from CBCs and other sources.
The SGWB from standard inflation~\cite{Grishchuk:1974ny,Starobinsky:1979ty,Grishchuk:1993te} ($\Omega_{\rm GW} \sim 10^{-15}$) is below the range of the plot. }
\label{fig_Omegas_final}
\end{figure}

Figure~\ref{fig_Omegas_final} shows the sum of the (minimized) $\Omega_{\rm unres}+\Omega_{\rm err}$ contributions from BBHs and BNSs for different waveform models and detector networks. The grey band shows the 90\% confidence level due to astrophysical uncertainties in our fiducial case ({\tt IMRPhenomD} and a 3-detector network).
We overplot the sensitivities of 3-detector and 5-detector networks to SGWBs, computed in the absence of foregrounds from CBCs and other sources, assuming the data is integrated for one year.  We also show the energy densities of SGWBs from other possible sources~\cite{Sachdev:2020bkk}, including: (i) axion inflation~\cite{Barnaby:2011qe}, (ii) post-inflation oscillations of a fluid with an equation of state stiffer than radiation~\cite{Boyle:2007zx}, (iii) a network of cosmic strings~\cite{Damour:2004kw, Siemens:2006yp, Olmez:2010bi, Regimbau:2011bm}, (iv) the most optimistic prediction from cosmic supernovae throughout the Universe~\cite{Finkel:2021zgf}, and (v) post-inflation preheating models aided by parametric resonance~\cite{Khlebnikov:1997di, Tilley:2000jh, Dufaux:2010cf, Figueroa:2017vfa}.  Note that the energy densities of these sources are model dependent, and they could be larger or smaller depending on the choice of model parameters.

In summary, subtracting the SGWB foreground from BBHs and BNSs is much harder than previously estimated.
As shown in Fig.~\ref{fig_freq_wise} for our fiducial case of a 3-detector network and {\tt IMRPhenomD}, the BBH background subtraction only reduces it by a factor of 2--3. Crucially, the remaining background still overwhelms the BNS background at frequencies below $\sim 100$~Hz. Similarly, the BNS background subtraction only reduces it by a factor of $\lesssim 2$.

\section{Conclusions}
\label{sec_concl}

The recent detections of nearly 100 individual CBC events (one with electromagnetic counterpart) have given us important information about astrophysics, cosmology, and fundamental physics.
Future observations of SGWBs, which are the superposition of many individually unresolvable events (either astrophysical or cosmological in nature), also carry invaluable physical information.
In the current (second-generation) detectors, BBHs and BNSs dominate, making it hard to observe astrophysical or cosmological SGWBs. More sensitive XG detectors will allow us to detect many more individual BBHs and BNSs, and to better measure their parameters. The question we address in this paper is whether these individual signals can be characterized and subtracted well enough to observe other, subdominant SGWBs.

We simulate BBH and BNS populations based on our current best estimates of their mass and redshift distributions and locally measured merger rates, making reasonable assumptions for the other parameters (see Table~\ref{tab_pop}). We find that the minimum $\Omega_{\rm unres} + \Omega_{\rm err}$ for BBHs or BNSs is reached at an optimal SNR threshold $\rm SNR_{thr}$, which can be taken to be frequency independent. For BBHs, we estimate an optimal $\rm SNR_{thr}$ of 10 for our fiducial waveform model {\tt IMRPhenomD} and 12 for {\tt IMRPhenomC} in our fiducial 3-detector network. For BNSs, we estimate an optimal $\rm SNR_{thr}$ of 20 for {\tt IMRPhenomD} and 42 for {\tt IMRPhenomC}.\footnote{In our optimistic 5-detector network and for {\tt IMRPhenomD}, the corresponding values are 6 for BBHs and 20 for BNSs.} The difference between the two waveform models, while important, is always much smaller than current astrophysical rate uncertainties, with the only exception of BBHs at high frequencies, where the energy density is low and details of the merger/ringdown waveform make a difference.
More realistic estimates should also include the NS-BH binary population. Note that the distinction between BBH, BNS and NS-BH systems is blurred by parameter estimation errors on the component masses, and for data analysis purposes it may be more practical to consider all three CBC populations at the same time.

One of our main findings is that the minimized value of $\Omega_{\rm unres} + \Omega_{\rm err}$ in XG detector networks is much larger than previous estimates. 
By computing the contribution to $\Omega_{\rm err}$ from each parameter, we can determine which parameters dominate $\Omega_{\rm err}$. 
We find that the uncertainty in the coalescence phase $\phi_c$ dominates for nearly all frequencies, due to its degeneracy with the polarization angle $\psi$. Another important contribution comes from the luminosity distance $D_L$, due to its degeneracy with the inclination angle $\cos\iota$. We find large differences in the contribution from $\mathcal{M}_z$ and $\eta$ between {\tt IMRPhenomD} and {\tt IMRPhenomC}. These differences reflect waveform modeling systematics to some extent, but they do not affect too much our estimate of $\Omega_{\rm err}$, which is comparable in the two models because $\phi_c$ dominates.

For our fiducial 3-detector network, the minimized $\Omega_{\rm unres} + \Omega_{\rm err}$ for BBHs is only a factor of 2--3 smaller than the total BBH SGWB, and it still overwhelms the background from BNSs below $\sim 100$~Hz. For BNSs, the minimized $\Omega_{\rm unres} + \Omega_{\rm err}$ is more than half of the total BNS background. 
As shown in Fig.~\ref{fig_Omegas_final}, even for the optimistic 5-detector network, the BBH and BNS subtraction can only be improved by a factor of $< 2$ compared to the 3-detector network.
This large residual makes it difficult to look for other SGWB sources.

In conclusion, subtracting the SGWB from BBHs and BNSs may be harder than anticipated. We hope that this will stimulate further work on data analysis techniques. 
The residuals due to imperfect removal could be reduced by subtracting the component tangent to the signal manifold at the point of best fit. This approach has been first proposed by Ref.~\cite{Cutler:2005qq, Sharma:2020btq}, although more detailed investigations are needed to understand the extent of this reduction when applied to realistic astrophysical catalogs.
Other possibilities include notching in the time-frequency plane~\cite{Zhong:2022ylh}, using Bayesian techniques to estimate the foreground and background signal parameters simultaneously~\cite{Biscoveanu:2020gds}, adopting a principal-component analysis to extract subdominant SGWBs~\cite{Pieroni:2020rob}, or exploiting the design topology of ET to construct a null stream~\cite{Freise:2008dk} that will help in understanding the foreground of CBC events~\cite{Regimbau:2012ir}.
One could also take advantage of the temporal and positional information of each event for a more precise subtraction.

The exceptional scientific payoff of a future detection of astrophysical and cosmological SGWBs motivates enduring efforts in the characterization of CBC SGWBs and further work on data analysis methods for their subtraction.

\section*{Acknowledgments}

We thank Sylvia Biscoveanu, Ssohrab Borhanian, Roberto Cotesta, Mark Hannam, and Alan Weinstein for helpful discussions. 
M.K. and B.Z. were supported by NSF Grant No.\ 2112699 and the Simons Foundation.
E.B., M.\c{C}. and L.R. are supported by NSF Grants No. AST-2006538, No. PHY-2207502, No. PHY-090003, and No. PHY20043, and NASA Grants No. 19-ATP19-0051, No. 20-LPS20- 0011, and No. 21-ATP21-0010. 
M.\c{C}.\ is also supported by Johns Hopkins University through the Rowland Research Fellowship. 
B.S.S. is supported by NSF Grants No. AST-2006384, No. PHY-2012083, and No. PHY-2207638. 
Part of E.B.'s and B.S.S.'s work was performed at the Aspen Center for Physics, which is supported by National Science Foundation Grant No. PHY-1607611. This research was supported in part by the National Science Foundation under Grant No. NSF PHY-1748958.
This research project was conducted using computational resources at the Maryland Advanced Research Computing Center (MARCC).
This work was carried out at the Advanced Research Computing at Hopkins (ARCH) core facility (\url{rockfish.jhu.edu}), which is supported by the NSF Grant No. OAC-1920103.
The authors acknowledge the Texas Advanced Computing Center (TACC) at The University of Texas at Austin for providing {HPC, visualization, database, or grid} resources that have contributed to the research results reported within this paper \cite{10.1145/3311790.3396656}. URL: \url{http://www.tacc.utexas.edu}.

\appendix

\section{Network sensitivity curves}
\label{app:sensitivities}

\begin{figure*}[t!]
\includegraphics[width=\columnwidth]{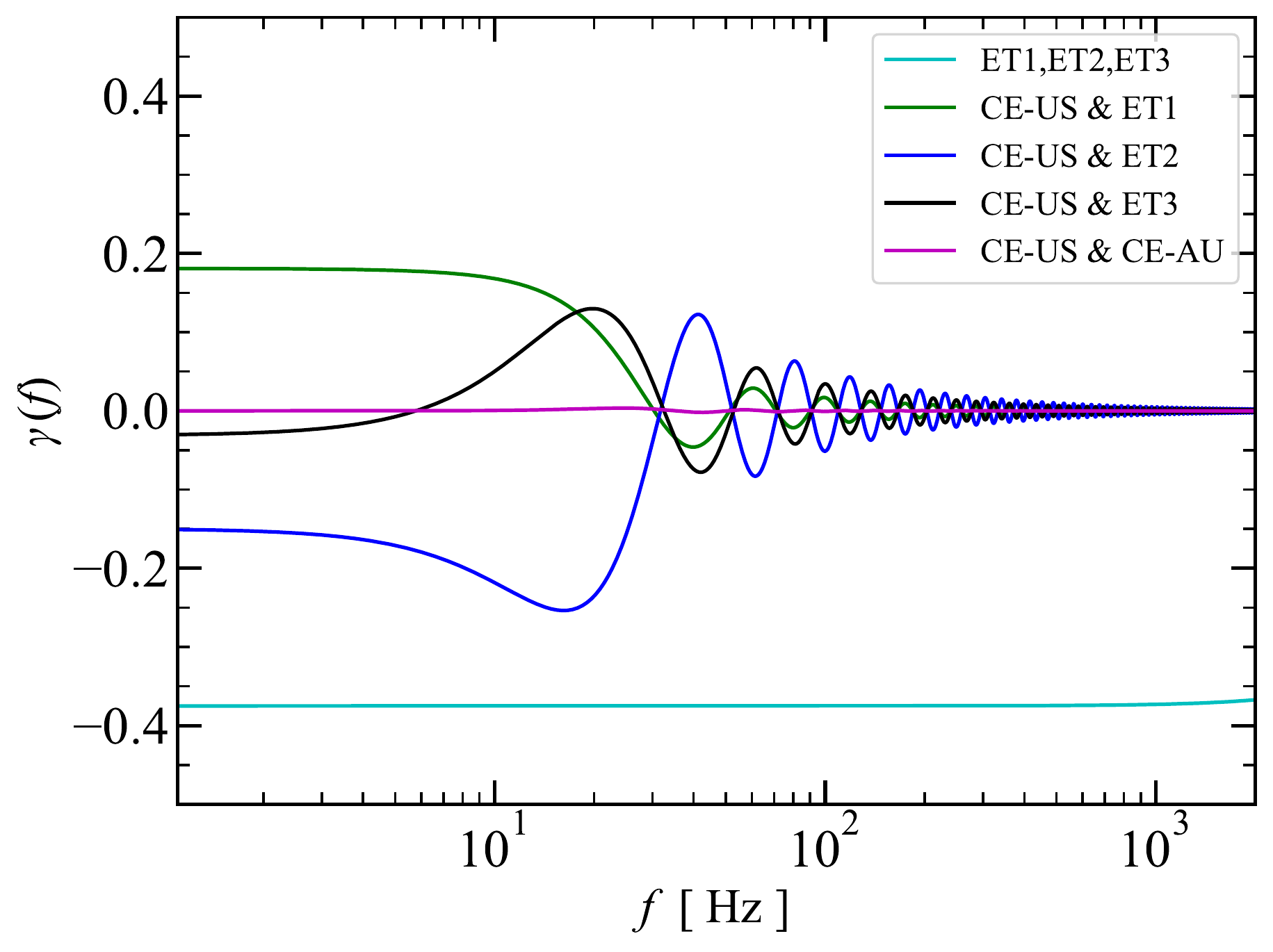}
\includegraphics[width=\columnwidth]{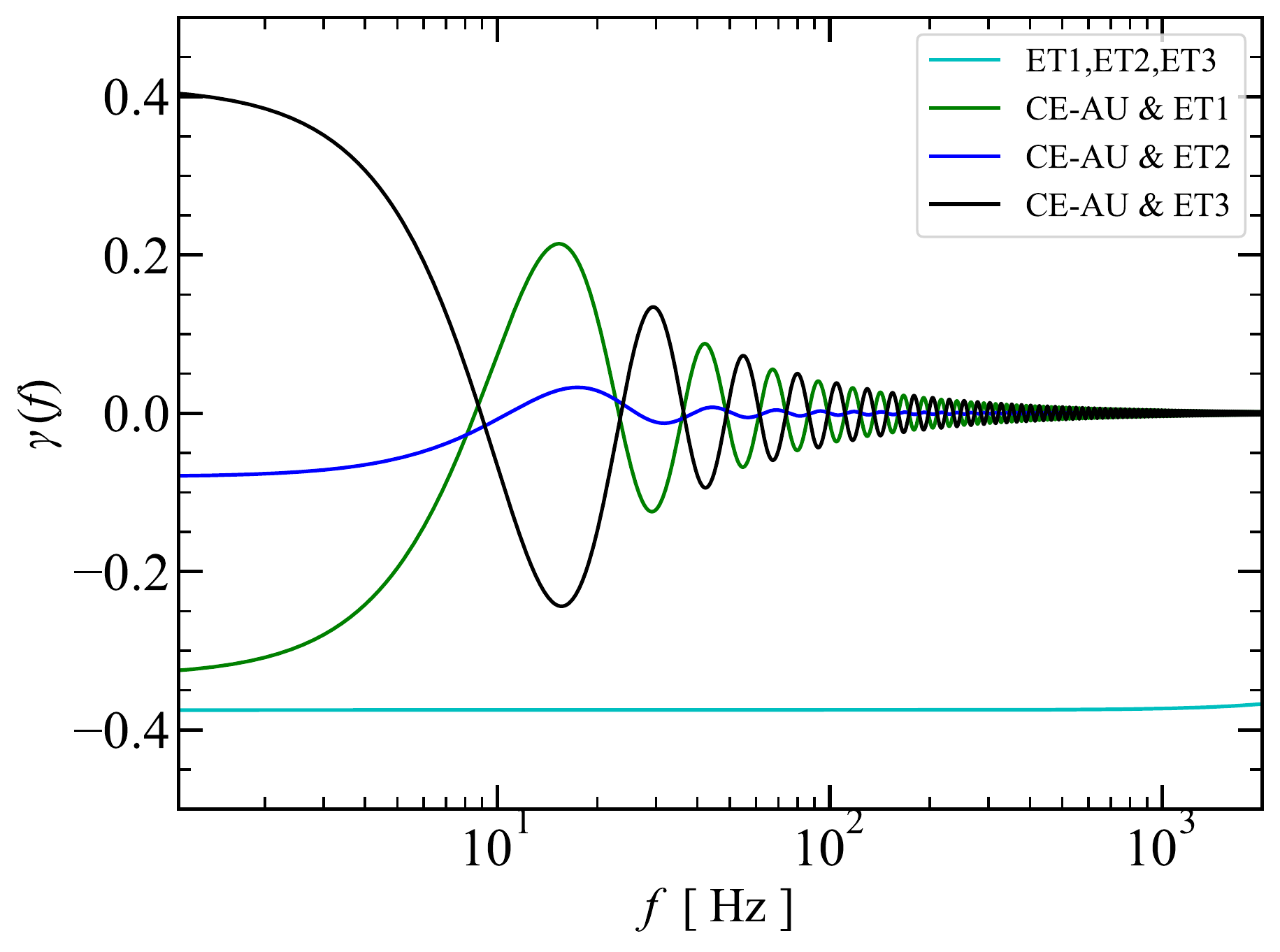}
\caption{ORFs for all the detector pairs in our fiducial network. The ORFs of the $40~\rm{km}$ CE detector located in the US with all other detectors in the network are shown in the left panel, while  the ORFs for the $20~\rm{km}$ CE in Australia are in the right panel. Both panels also show the ORF for any detector pair in the ET triangular configuration (cyan line), which is approximately constant in the frequency range considered here. The amplitude of the ORF for the CE pair is close to zero due to the relative orientations of the detectors.}
\label{fig_orf}
\end{figure*}

In this Appendix, we briefly summarize the formalism adopted to estimate the network sensitivity curve shown as a reference in Fig.~\ref{fig_Omega_plus_uncert}. We also compute the overlap reduction functions (ORFs) for the fiducial locations and orientations of the CE detectors adopted in \texttt{GWBENCH}~\cite{Borhanian:2020ypi}. We mostly follow Ref.~\cite{Thrane:2013oya} for the definition of sensitivity curves, and Refs.~\cite{Allen:1997ad,Flanagan:1993ix} for the calculation of the ORFs.

Searches for SGWBs in instrumental noise are conducted by cross-correlating the outputs of multiple GW detectors (see e.g.~\cite{KAGRA:2021kbb}). The optimal SNR for an unpolarized, isotropic SGWB in a network of $N_{\rm det}$ interferometers with PSDs $S_{n,i}(f),S_{n,j}(f)$ is given by~\cite{Maggiore:1999vm,Thrane:2013oya}
\begin{equation}
    {\rm SNR_{b}} =\frac{3 H_0^2}{10 \pi^2}  \left[ 2T
\int_0^\infty df\>
\sum_{i=1}^{N_{\rm det}}\sum_{j>i}^{N_{\rm det}}
\frac{\gamma_{ij}^2(f)\Omega_{\rm GW}^2(f)}{f^6 S_{n,i}(f)S_{n,j}(f)} \right]^{1/2}.
\label{eq_snr_back}
\end{equation}
Here $T$ is the observing time and $\gamma_{ij}(f)$ is the ORF~\cite{Allen:1997ad,Flanagan:1993ix}, a dimensionless function of frequency that accounts for the reduction in sensitivity due to the different positions and orientations of the detectors in the network. Explicitly, the ORF between the $i$th and $j$th detector in the network is defined as~\cite{Allen:1997ad,Flanagan:1993ix}:
\begin{equation}
    \gamma_{ij}(f) = \frac{5}{8\pi}\sum_A\int_{S^2}d\hat{\Omega}\, F_i^A(\hat{\Omega})F_j^A(\hat{\Omega})e^{i 2\pi\hat{\Omega}\cdot\vec{\Delta x}} \,,
    \label{eq_orf}
\end{equation}
where $\hat{\Omega}$ is a unit vector specifying the direction on a sphere, $\vec{\Delta x}$ the separation vector between the two detector sites, and $F_i^A(\hat{\Omega}),F_j^A(\hat{\Omega})$ are the response functions of the two detectors $i, j$ to the $A=+,\times$ polarizations. The normalization factor $5/8\pi$ is introduced so that $\gamma_{ij}(f)=1$ for colocated and cooriented interferometers~\cite{Allen:1997ad}. A closed form for the integral~\eqref{eq_orf} is derived in Refs.~\cite{Allen:1997ad,Flanagan:1993ix}.

Figure~\ref{fig_orf} shows the ORFs for our fiducial 3-detector network. The ORFs for detector pairs in the ET triangle are approximately constant, with a value of about $\gamma_{ij}(f)\approx -0.38$ in the entire observation band~\cite{Sharma:2020btq,Amalberti:2021kzh}, as the detectors are almost colocated. The amplitude of the ORFs for the CE detectors with any detector in the network drops rapidly, meaning that the network sensitivity to SGWBs will be dominated by ET at high frequency. The amplitude of the ORF for a pair of CEs located in the U.S. and in Australia is close to zero due to the relative orientations of the interferometers: in our fiducial setting, the projection of the bisectors of these two detectors onto the local horizontal plane are rotated by an angle of about $\pi/4$ with respect to each other~\cite{Borhanian:2020ypi}.

By analogy with Eq.~\eqref{eq_snr}, we can define an effective network noise PSD~\cite{Thrane:2013oya}
\begin{equation}
    S_{n,\rm{net}}(f) = \left[
\sum_{i=1}^{N_{\rm det}}\sum_{j>i}^{N_{\rm det}}
\frac{\gamma_{ij}^2(f)}{S_{n,i}(f)S_{n,j}(f)} \right]^{-1/2}\,.
\end{equation}
Any noise PSD $S_n(f)$ can be interpreted as the mean square amplitude of the noise per unit frequency, and is associated to a spectral energy density $\propto f S_n(f)$~\cite{Moore:2014lga,Schutz:2011tw}. Therefore, by applying the definition [Eq.~\eqref{eq_omegadef}], we can introduce an \emph{effective} dimensionless noise energy spectrum for the detector network~\cite{Thrane:2013oya,Mingarelli:2019mvk}
\begin{equation}
    \Omega_n(f) = \frac{10 \pi^2}{3 H_0^2}\,f^3 S_{n,\rm{net}}(f)\,,
\end{equation}
such that the optimal SNR [Eq.~\eqref{eq_snr_back}] can be rewritten as
\begin{equation}
    {\rm SNR_{b}} =  \left[ 2T
\int_0^\infty df\>
\frac{\Omega_{\rm GW}^2(f)}{\Omega_n^2(f)} \right]^{1/2}.
\label{eq_snr_back_net}
\end{equation}

Most SGWBs take the form of power laws, i.e., 
\begin{equation}
    \Omega_{\rm GW}(f) = \Omega_\beta\left(\frac{f}{f_{\rm ref}}\right)^\beta\,,
    \label{eq_powerlaw_sgwb}
\end{equation}
where $f_{\rm ref}$ is an arbitrary reference frequency and $\Omega_\beta$ is the amplitude of the background evaluated at $f=f_{\rm ref}$. Most cosmological backgrounds are pure power laws~\cite{Barnaby:2011qe,Damour:2004kw,Siemens:2006yp,Boyle:2007zx}, and even the CBC background can be approximated by a power law at lower frequencies, in the regime where the inspiral phase dominates~\cite{Maggiore:1999vm}.
By plugging Eq.~\eqref{eq_powerlaw_sgwb} into Eq.~\eqref{eq_snr_back_net}, we can compute the amplitude that corresponds to ${\rm SNR_b}=1$ for a given spectral index $\beta$:
\begin{equation}
    \Omega_{\beta} = \left[ \frac{1}{2T}
\int_0^\infty df\>
\frac{(f/f_{\rm ref})^{2\beta}}{\Omega_n^2(f)} \right]^{-1/2} \,.
\end{equation}
Each pair $(\Omega_{\beta},\beta)$ corresponds to a different power-law background~\eqref{eq_powerlaw_sgwb} with network SNR equal to one. We call the \emph{power-law integrated sensitivity curve} the envelope of all these backgrounds, i.e.~\cite{Thrane:2013oya}
\begin{equation}
    \Omega_{\rm PI}(f) = \max_\beta\left[ \Omega_\beta\left(\frac{f}{f_{\rm ref}}\right)^\beta \right] \,.
\end{equation}
This curve represents a natural way to estimate the detector network sensitivity to SGWBs, as any power-law SGWB that lies somewhere above the curve has ${\rm SNR_b} > 1$, and thus it may be detectable.

The network sensitivity shown in Fig.~\ref{fig_Omega_plus_uncert} is a power-law integrated sensitivity curve computed with the formalism described here for our fiducial 3-detector network, assuming an observing time $T=1~\rm{yr}$.

\bibliography{GWB}

\end{document}